\documentclass{aa}  

\usepackage{graphicx}
\usepackage{txfonts}
\usepackage[pdfpagelabels=false]{hyperref}
\hypersetup{colorlinks=true,linkcolor=blue,citecolor=blue,filecolor=blue,urlcolor=blue,}
\makeatletter
\renewcommand*\aa@pageof{, page \thepage{} of \pageref*{LastPage}}
\makeatother

\usepackage{graphicx}	% Including figure files
\usepackage{amsmath}	% Advanced maths commands
\usepackage{amssymb}	% Extra maths symbols
\usepackage{multirow}   % Extra command \multirow
\usepackage{array}      % 
\usepackage{mathtools}

\usepackage{xcolor}
\definecolor{green_comm}{RGB}{0,160,0}

\graphicspath{{./Figures/}}

\newcolumntype{P}[1]{>{\centering\arraybackslash}p{#1}}

\newcommand{\de}{\partial}
\newcommand{\Hy}{\ion{H}{}}
\newcommand{\HI}{\ion{H}{i}}
\newcommand{\HII}{\ion{H}{ii}}
\newcommand{\He}{\ion{He}{}}
\newcommand{\HeI}{\ion{He}{i}}
\newcommand{\HeII}{\ion{He}{ii}}
\newcommand{\HeIII}{\ion{He}{iii}}
\newcommand{\der}[2]{\displaystyle \frac{\de #1}{\de #2}}
\newcommand{\mdot}{$\dot{M}$}

\newcommand{\gtsima}{$\; \buildrel > \over \sim \;$}
\newcommand{\ltsima}{$\; \buildrel < \over \sim \;$}
\newcommand{\prosima}{$\; \buildrel \propto \over \sim \;$}
\newcommand{\gsim}{\lower.5ex\hbox{\gtsima}}
\newcommand{\lsim}{\lower.5ex\hbox{\ltsima}}
\newcommand{\simgt}{\lower.5ex\hbox{\gtsima}}
\newcommand{\simlt}{\lower.5ex\hbox{\ltsima}}
\newcommand{\simpr}{\lower.5ex\hbox{\prosima}}

\newcommand\Tstrut{\rule{0pt}{2.6ex}}         % = `top' strut
\newcommand\Bstrut{\rule[-0.9ex]{0pt}{0pt}}   % = `bottom' strut

\begin{document}

\title{Irradiation-driven escape of primordial planetary atmospheres I. The ATES photoionization hydrodynamics code\thanks{The code is publicly available at \url{https://github.com/AndreaCaldiroli/ATES-Code}.}}

   %\subtitle{ }

   \author{Andrea Caldiroli\inst{1,2}
          \and Francesco Haardt\inst{1,3,4}
          \and Elena Gallo\inst{5,6} 
          \and Riccardo Spinelli\inst{1,4}
          \and Isaac Malsky\inst{5}
          \and Emily Rauscher\inst{5}}

   \institute{Dipartimento di Scienza e Alta Tecnologia, 
   			  Universit\`a degli Studi dell'Insubria, 
   			  via Valleggio 11, 22100 Como, Italy
        \and
             Fakult\"at f\"ur Mathematik, Universit\"at Wien, Oskar-Morgenstern-Platz 1, A-1090 Wien, Austria
       \and
             INFN -- Sezione Milano-Bicocca, Piazza della Scienza~3, 20126 Milano, Italy
         \and
              INAF -- Osservatorio Astronomico di Brera, Via E. Bianchi 46, 23807 Merate, Italy
          \and
         	 Department of Astronomy, University of Michigan, 1085 S University, Ann Arbor, Michigan 48109, USA
         \and
             Dipartimento di Fisica ``G. Occhialini", Universit\`a degli Studi di Milano-Bicocca, Piazza della Scienza~3, 20126 Milano, Italy   
             }

%\date{Received September 15, 1996; accepted March 16, 1997}

\abstract{
Intense X-ray and ultraviolet stellar irradiation can heat and inflate the atmospheres of closely orbiting exoplanets, driving mass outflows that may be significant enough to evaporate a sizable fraction of the planet atmosphere over the system lifetime. The recent surge in the number of known exoplanets, together with the imminent deployment of new ground and space-based facilities for exoplanet discovery and characterization, requires a prompt and efficient assessment of the most promising targets for intensive spectroscopic follow-ups. 
To this purpose, we developed ATES (ATmospheric EScape); a new hydrodynamics code that is specifically designed to compute the temperature, density, velocity and ionization fraction profiles of highly irradiated planetary atmospheres, along with the current, steady-state mass loss rate.
ATES solves the one-dimensional Euler, mass and energy conservation equations in radial coordinates through a finite-volume scheme. The hydrodynamics module is paired with a photoionization equilibrium solver that includes cooling via bremsstrahlung, recombination and collisional excitation/ionization for the case of a primordial atmosphere entirely composed of atomic hydrogen and helium, whilst also accounting for advection of the different ion species.   
Compared against the results of 14 moderately-to-highly irradiated planets simulated with The PLUTO-CLOUDY Interface, which couples two sophisticated and computationally expensive hydrodynamics and radiation codes of much broader astrophysical applicability, ATES yields remarkably good agreement at a significantly smaller fraction of the time. 
A convergence study shows that ATES recovers stable, steady-state hydrodynamic solutions for systems with $\log(-\phi_p) \lesssim 12.9 + 0.17\log F_{\rm XUV}$, where $\phi_p$ and $F_{\rm XUV}$ are the planet gravitational potential and stellar flux (in cgs units). Incidentally, atmospheres of systems above this threshold are generally thought to be undergoing Jeans escape. The code, which also features a user-friendly graphic interface, is available publicly as an online repository.}
\keywords{Planets and satellites: atmospheres -- Planets and satellites: dynamical evolution and stability -- Hydrodynamics --  Methods: numerical}

\titlerunning{ATES}
\maketitle
%
%-------------------------------------------------------------------

\section{Introduction}
%\er{feel free}~\im{to edit}
Atmospheric stability conditions of gaseous planets in close proximity to their host star has been matter of investigation since the first observations of exoplanets in 1995 \citep{Mayor1995}. Exposure to intense ultraviolet (UV) and X-ray irradiation is bound to cause physical and chemical atmospheric evolution by dissociating molecules above the planet radius and producing a mixture of neutral and ionized atoms at higher altitudes \citep{Vidal-Madjar2003,Yelle2004,Tian2005,Koskinen2014}. The integrated amount of energy that is absorbed by the atmosphere of a close-in planet over its lifetime could amount to a sizable fraction of its gravitational binding energy, yielding hot (with temperatures in the range $5,000-10,000$~K), weakly bounded thermospheres that may be prone to substantial evaporation. Ongoing atmospheric escape has confirmed observationally in a handful of nearby exoplanets through the detection of escaping hydrogen via transit Ly$\alpha$ spectroscopy (e.g., HD~209458~b, \citealt{Vidal-Madjar2003}; HD~189733~b, \citealt{Lecavelier2010,Bourrier2013}; GJ~436~b, \citealt{Kulow2014,Ehrenreich2015}) and, more recently, through the detection of helium in the outer atmosphere of the sub-Saturn WASP-107~b \citep{Spake2018}.

\citet{Watson1981} derived the first analytical expression for atmospheric mass loss rate, based on the assumption that the incident stellar radiation is partially converted into expansion work. The instantaneous rate of mass loss $\dot M$ from an irradiated atmosphere is expected to depend directly upon the incident UV-to-X-ray flux and inversely upon the planetary density. It is then predicted that gas giants in close orbits around UV/X-ray luminous stars should experience strong irradiation and consequent high rates of mass loss, leading to the removal of a substantial portion of their initial light element gas envelope.
In the context of the so-called ``energy-limited'' approximation, however, the heating efficiency can not be readily estimated from first principles (see \citealt{Krenn2021} for detailed discussions). Radiative losses, which are unaccounted for in this formulation, cannot be neglected in a high-irradiation regime where the recombination timescale is shorter than the outflow dynamical timescale \citep{Lammer2003}. 
In the case where radiative cooling dominates over adiabatic expansion, such as, e.g., for hot Jupiters, the energy-limited formalism can over-estimate the actual mass outflow rates by orders of magnitude (see \citealt{Owen2019}, and references therein). 

These and other studies indicate that, although hydrodynamic escape should have modest effects in reducing the mass of hot Jupiters, it could play a significant role in shaping the observed properties of the known (hot) exoplanet population, likely contributing to carving the observed radius valley by stripping planets of their H/He atmospheres and turning them into remnant rocky cores \citep{lammer09,Eh11,lopez12,owenwu13,owenwu17,fulton17,Jin2018,kuby20}. 

A full understanding of photoevaporative loss is thus warranted for deciphering the full picture of planet formation and evolution; as a result, in the last two decades, much effort has gone into the development of more realistic, numerical models of atmospheric escape \citep{Lammer2003,Yelle2004,Tian2005,GarciaMunoz2007,MurrayClay2009,Owen2012,Erkaev2013,Erkaev2015,Erkaev2016,Salz2015,Debrecht2019,McCann2019,Esquivel2019,Vidotto2020}. Specific exoplanet targets have been modeled with a great deal of sophistication, also accounting for 2D and/or 3D effects and complex chemistry (see, e.g, \citealt{gj} and \citealt{Khodachenko2019} for GJ~436~b, \citealt{koskinen13}, \citealt{Khodachenko2017}, \citealt{Bisikalo2018} and \citealt{Debrecht2020} for HD~209458~b, \citealt{odert20} for HD~189733~b). 
Typically, these models aim to reproduce the results of time-intensive transit spectroscopy campaigns, and are extremely computationally expensive; as a corollary, the associated numerical solvers are seldom publicly available. A complementary approach consists of generating extensive grids of hydrodynamical models with fixed heating efficiency, covering a wide range of planetary masses and stellar parameters, which can then be interpolated to best approximate the planet of choice \citep{kuby18a,kuby18b,kuby21}.\\

The next decade will usher a new generation of visible and infrared instrumentation for the detection and characterization of exoplanets. This ever-growing parameter space demands rapid and reliable estimates of the expected atmospheric parameters for targeted follow-ups. 
To this end, we developed ATES (ATmospheric EScape); a new, open-source, user-friendly hydrodynamics code designed to compute the temperature, density, velocity and ionization fraction profiles of strongly irradiated, primordial planetary atmospheres composed of atomic hydrogen and helium.

In this Paper--the first in a series of three--we describe, test and validate ATES by comparing our results to those of 
TPCI (The PLUTO-CLOUDY Interface; \citealt{Salz2015}), a publicly available interface between the magneto-hydrodynamics code PLUTO \citep{pluto} and the plasma simulation and spectral synthesis code CLOUDY \citep{cloudy}, applied to the specific case of irradiated planetary atmospheres (\citealt{Salz2016a}; hereafter S16). The combined effect of planetary gravity and stellar irradiation intensity on the thermal escape hydrodynamics will be explored in Paper II, whereas Paper III will present revised estimates of mass outflow rates for a distance-limited exoplanet sample, with planetary parameters updated based on revised parallactic distances, from Gaia. \\

The code hydrodynamics and radiation modules are described in \S\ref{sec:ModelDescription}; \S\ref{sec:NumericalMethods} details the numerical solvers for the Euler and radiative equilibrium equations; \S\ref{sec:Validation} tests the code by simulating two standard astrophysical problems with well known analytical solutions, whereas \S\ref{sec:Setup} describes the recommended code setup for the purpose of simulating escaping planetary atmospheres, along with possible choices of different numerical routines. 
In \S\ref{sec:Results}, we present and discuss our results vis-à-vis those from TPCI for a sample of 14 nearby exoplanets (S16), perform a detailed convergence study, and lay out future possible applications of and addition to ATES. 
%--------------------------------------------------------------------

\section{Model}
\label{sec:ModelDescription}

\subsection{Hydrodynamics}
\label{subsec:ModelHydro}
The dynamics of the atmospheric gas is described by Euler equations in the presence of a gravitational field. Under the assumption of spherical symmetry, Euler equations can be written in a conservative, one dimensional form as:
%%%%%%%%%%%%%%%%%%%%%%%%%%%%%%
\begin{equation}
    \begin{cases}
        \der{\rho}{t} + \frac{1}{r^2}\der{\rho v r^2}{r} = 0 \\[8pt]
        \der{\rho v}{t} + \frac{1}{r^2}\der{\rho v^2r^2}{r} + \der{p}{r} = -\rho \der{\Phi}{r} \\[8pt]
        \der{E}{t} + \frac{1}{r^2}\der{(E + p)vr^2}{r} = -\rho v\der{\Phi}{r} + Q
    \end{cases}
\label{eq:EulerEquations}
\end{equation}
%%%%%%%%%%%%%%%%%%%%%%%%%%%%%%
where $r$ is the distance from the planet center, $\rho$, $v$ and $p$ are the mass density, velocity and pressure of the outflowing gas; $E = \rho v^2/2+p/(\gamma -1)$ is the gas total energy density (where $\gamma = 5/3$). The function $Q(r)$ accounts for the heating and cooling mechanisms; its expression will be detailed in \S\ref{subsec:RadiativeMechanisms}. 

We adopt the following expression for the gravitational potential $\Phi$ \citep[e.g.][]{Erkaev2007}:
%%%%%%%%%%%%%%%%%%%%%%%%%%%%%%
\begin{equation}
    \Phi = -\frac{GM_p}{r}-\frac{GM_\star}{a-r}-\frac{G(M_p+M_\star)}{2a^3}\left(r-a\frac{M_\star}{M_p+M_\star}\right)^2,
\end{equation}
where $G$ is the gravitational constant, $M_p$ the planet mass, $M_\star$ the star mass and $a$ the (average) orbital distance. This above expression also accounts for effects of the Roche potential, which may significantly affect the inferred atmospheric mass loss rate \citep[see, e.g.,][]{Lecavalier2004,Jaritz2005,Erkaev2007,kuby18a}.

We assume a primordial atmosphere composed of atomic hydrogen  (neglecting all forms of molecular hydrogen, which should not become dominant until deeper in the atmosphere) and helium, with relative abundances set to \He/\Hy=0.083 in number density (the relative abundance value can be set by the user in the publicly available version of ATES). ATES only treats ion interactions through electron collisions; ion-ion interactions are not included, implying that there is no direct energy exchange between the different species.
We stress that our solution scheme (see next Section) does not allow for radial mixing amongst the different elements, in the sense that, while the ionization state of the gas can change as a function of the distance from the planet, the overall He/H ratio is kept constant throughout the simulation. Finally, we adopt the equation of state of an ideal gas, where $p = \sum n_i k_B T$, where $T$ is the gas temperature, $k_B$ the Boltzmann's constant, and the summation is meant over all the species (including free electrons), each with number density $n_i$.
%%%%%%%%%%%%%%%%%%%%%%%%%%%

\subsection{Energy and ionization balance}
\label{subsec:RadiativeMechanisms}
The source function $Q(r)$ in Equation~\ref{eq:EulerEquations} represents the net energy deposition at coordinate $r$; it is written as $Q=\mathcal{H}-n_e\Lambda$, where $\mathcal{H}$ and $\Lambda$ are the total heating and cooling rates, respectively, and $n_e$ is the total number density of free electrons.

Heating of the thermosphere is provided by the stellar photoionizing radiation (photo-heating). For a given stellar luminosity $L_E$, the photo-heating rate at height $r$ is given by:
%%%%%%%%%%%%%%%%%%%%%%%%%%%%%%
\begin{equation}
	\mathcal{H}(r) = \sum_{i}n_i\int\limits_{E_{t,i}}^\infty{dE\, F_E\left(1-\frac{E_{t,i}}{E}\right)e^{-\tau_E}\sigma_{E,i}},
\label{eq:HeatingRate1D}
\end{equation}
%%%%%%%%%%%%%%%%%%%%%%%%%%%%%%
where we set the average photo-heating flux seen by the planet equal to $F_E=L_E/(4\pi a^2)$, as $a$ is typically much larger than the hydrostatic pressure scale-height.
The subscript $i$ refers to $\HI$ (for neutral hydrogen), $\HII$ (for ionized hydrogen), $\HeI$ (for neutral helium), $\HeII$ (for single ionized helium), and $\HeIII$ (for fully ionized helium), with $E_{t,i}$ and $\sigma_{E,i}$ representing the appropriate ionization thresholds and photoionization cross sections, respectively. The optical depth at $r$ is given by:
%%%%%%%%%%%%%%%%%%%%%%%%%%%%%%
\begin{equation}
	\tau_E (r) =\sum\limits_{i}{~\int\limits_{r}^{a}{dr' \, n_i(r')\sigma_{E,i}}}.
\label{eq:OpticalDepth}
\end{equation}
%%%%%%%%%%%%%%%%%%%%%%%%%%%%%%
%
Radiative cooling includes bremsstrahlung, recombination, collisional excitation, and collisional ionization. We adopt the rates given by \citet{Hui1997} and \citet{Glover2007}, and detailed in Appendix~\ref{app:rates}. The Euler equations include two additional heat transport terms, namely adiabatic expansion (cooling) and advection (heating and/or cooling), respectively proportional to $\propto p\de_r(vr^2)$ and $\propto \de_r(pvr^2)$.
%; the latter account for the radial transport of different ion species

Ion abundances are derived under the assumption of photoionization equilibrium. The steady-state ionization profiles can be obtained by solving the following system of equations:
%%%%%%%%%%%%%%%%%%%%%%%%%%%%%%
\begin{equation}
	\begin{cases}
	\begin{aligned}
	\frac{1}{r^2}\der{n_{\HII} v r^2}{r}	 =\Gamma_{\HI}~n_\HI + (\alpha^{\text{ion}}_\HI~n_\HI -\alpha^{\text{rec}}_\HII~ n_\HII)~n_e 
	\end{aligned}\\
	\begin{aligned}
	\frac{1}{r^2}\der{n_{\HeII} v r^2}{r}	& = \Gamma_{\HeI}~n_\HeI + (\alpha^{\text{ion}}_\HeI~n_\HeI -\alpha^{\text{rec}}_\HeII ~n_\HeII)~n_e \\
	& - \Gamma_{\HeII}~n_\HeII - (\alpha^{\text{ion}}_\HeII~n_\HeII -\alpha^{\text{rec}}_\HeIII~n_\HeIII)~n_e
	\end{aligned}\\
	\begin{aligned}
	\frac{1}{r^2}\der{n_{\HeIII} v r^2}{r}	= \Gamma_{\HeII}~n_\HeII + (\alpha^{\text{ion}}_\HeII~n_\HeII -\alpha^{\text{rec}}_\HeIII~n_\HeIII)~n_e,
	\end{aligned}
	\end{cases}
\label{eq:RadiativeEquilibriumSystem1}
\end{equation}
%%%%%%%%%%%%%%%%%%%%%%%%%%%%%%
where $\alpha^{\text{ion}}_i(T)$ and $\alpha^{\text{rec}}_i(T)$ are the collisional ionization and recombination coefficients (cm$^3$s$^{-1}$), respectively (see Appendix~\ref{app:rates} for numerical values), while $\Gamma_i$ is the photoionization rate per ion (s$^{-1}$): 
%%%%%%%%%%%%%%%%%%%%%%%%%%%%%%
\begin{equation}
	\Gamma_i = \int\limits_{E_{t,i}}^\infty{dE\, \frac{F_E}{E} e^{-\tau_E} \sigma_{E,i} }.
\label{eq:PhotoionizationRate1D}
\end{equation}
Recognizing that, in steady-state, the mass outflow rate $\dot{M} \propto (n_{\rm H}+n_{\rm He}) vr^2$ is constant with radius, the photoionization balance equations can be re-cast in terms of the different ion fractions $f_i$: 
%%%%%%%%%%%%%%%%%%%%%%%%%%%%%%
\begin{equation}
	\begin{cases}
	v \der{f_{\HII}}{r}	 = \Gamma_{\HI}~f_\HI + (\alpha^{\text{ion}}_\HI~f_\HI -\alpha^{\text{rec}}_\HII~f_\HII)~n_e \\
	\begin{aligned}
		v \der{f_{\HeII}}{r}		&  =\Gamma_{\HeI}~f_\HeI + (\alpha^{\text{ion}}_\HeI~f_\HeI -\alpha^{\text{rec}}_\HeII~ f_\HeII)~n_e \\
	& - \Gamma_{\HeII}~f_\HeII - (\alpha^{\text{ion}}_\HeII~f_\HeII -\alpha^{\text{rec}}_\HeIII~f_\HeIII)~n_e
	\end{aligned}\\
	v \der{f_{\HeIII}}{r}	= \Gamma_{\HeII}~f_\HeII + (\alpha^{\text{ion}}_\HeII~f_\HeII -\alpha^{\text{rec}}_\HeIII~f_\HeIII)~n_e.
	\end{cases}
\label{eq:RadiativeEquilibriumSystem2}
\end{equation}
%%%%%%%%%%%%%%%%%%%%%%%%%%%%%%
The above equations are complemented by charge conservation: $n_e = n_\HII + n_\HeII + 2n_\HeIII$; additionally, the helium-to-hydrogen ratio $n_{\rm He}/n_{\rm H}$ (whose value can be chosen by the user) is kept constant with radius. 
\subsection{Ion advection}
\label{sec:adv}
In principle, the photoionization equations above ought to be solved in tandem with Euler equations at each time step. 
Instead, ATES accounts for the role of ion advection--i.e., the radial transport of different ion species--\textit{in post-processing}, as follows.  
First, the photoionization balance equations are solved at each time-step under the assumption of stationary-state conditions, i.e., in the simplified case where the partial radial derivatives on the left-hand side of Equations~\ref{eq:RadiativeEquilibriumSystem2} are all set to zero; this amounts to neglecting advection altogether. 

In terms of dynamical quantities, i.e., velocity, density and pressure, this approach yields very good agreement with the profiles obtained through a dedicated CLOUDY module (see \S\ref{sec:Results}), where the photoionization equilibrium equations are solved concurrently with the hydrodynamics module at each time-step (S16). In contrast, the ionization profiles, which are especially sensitive to the effects of advection, are poorly recovered ; albeit to a lesser extent, the same is true for the temperature, since $p=(\rho/\mu)k_{B}T$, where $\mu$ is the mean molecular weight\footnote{The mean molecular weight can be expressed in terms of ionization fractions as:
%%%%%%%%%%%%%%%%%%%%%
\begin{equation*}
    \mu=\frac{f_\HI+f_\HII+4Y(f_\HeI+f_\HeII+f_\HeIII)}{f_\HI+2f_\HII+4Y(f_\HeI+2f_\HeII+3f_\HeIII)}\,m_u,
\end{equation*}
%%%%%%%%%%%%%%%%%%%%%
where $Y\equiv n_\He/n_\Hy$, and $m_u$ is the atomic mass unit.}.

As a next step, in place of solving the full system of non-linear transcendental integro-differential equations in $f_i$, ATES solves  Equations~\ref{eq:RadiativeEquilibriumSystem2} by adopting the stationary-ionization solutions as Ansatz for $\rho(r)$, $v(r)$, and $T(r)$. 
%the density, velocity and temperature profiles.
Specifically, the temperature profile is used to estimate directly the values of ${\alpha^{\text{ion}}_i}$ and $\alpha^{\text{rec}}_{i}$, wheres the density profile is adopted to approximate $\Gamma_i$. 
This also enables the solver to bypass knowledge of the outer boundary conditions that would be necessary to properly solve Equation~\ref{eq:OpticalDepth} and thus evaluate $\Gamma_i$; with this approach, an \textit{inner} boundary condition can be set instead, by imposing full neutrality ($f_\HI=f_\HeI=1$) at the planet radius, i.e., the radius at which the planet becomes optically thick to visible light.  

\begin{figure}
\centering
    \includegraphics[width=0.9\columnwidth]{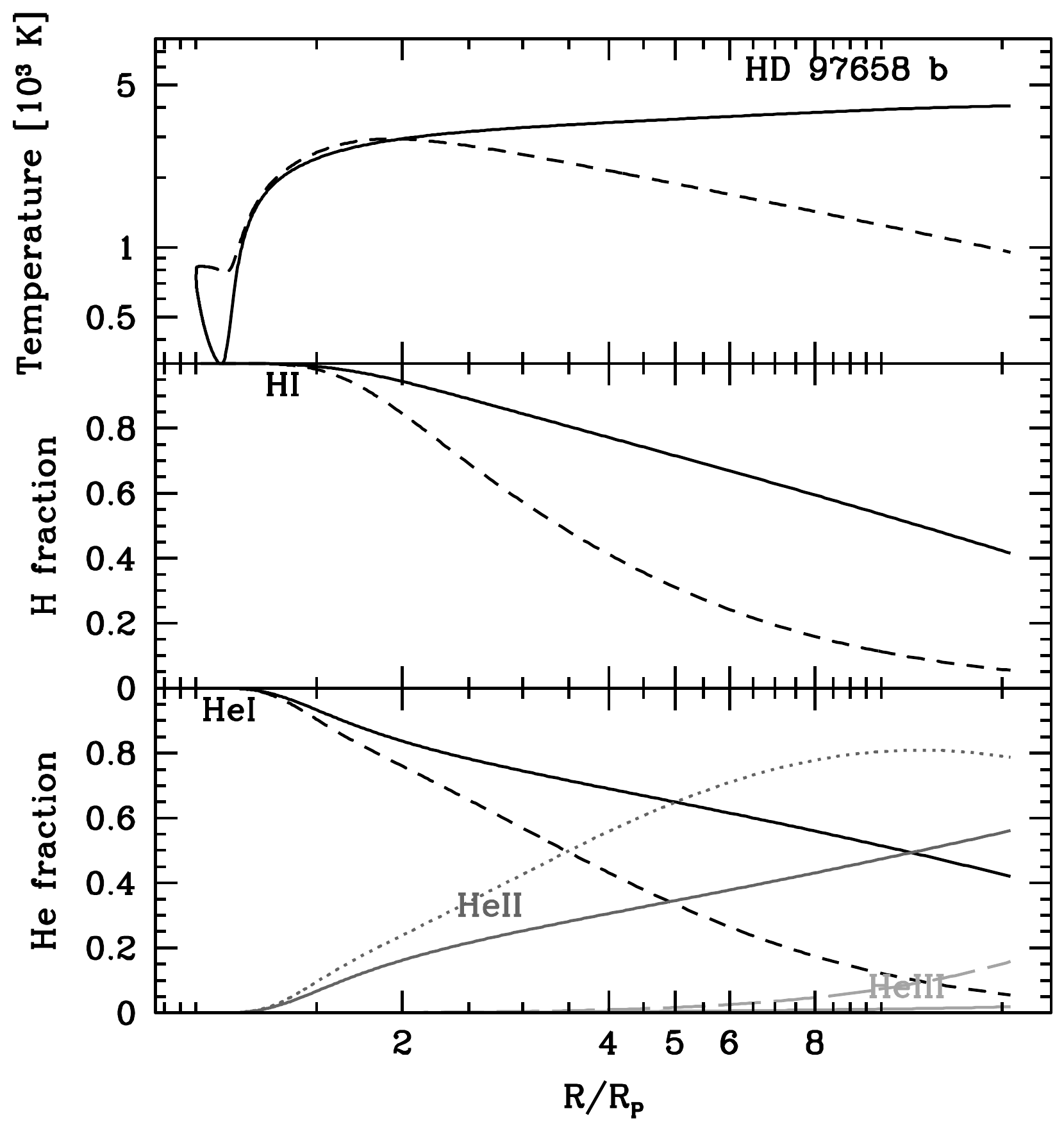}
    %\vspace{-3cm}
    \caption{The effect of ion advection on the temperature (top) hydrogen (middle) and helium (bottom) profiles are shown here for the case of HD~97658~b. The dashed/dotted lines trace the profiles calculated neglecting the role advection, i.e. by solving the photoionization equilibrium equations in tandem with Euler equations, assuming stationary conditions; the solid lines illustrate how the profiles change when advection is implemented in post-processing, as described in \S\ref{sec:adv}. The relatively shallow ionization front in HD~97658~b renders this effect more extreme compared to other planets. 
    %The results are in good agreement with those obtained by TPCI (S16), shown as dashed red lines in Figure~\ref{fig:hd-six} for the same planet.
    }
    \label{fig:IonAdvection}
\end{figure}

As shown in Figure~\ref{fig:IonAdvection}, the resulting ionization profiles exhibit fairly large differences compared to those that are obtained assuming stationary conditions. However, this approach yields a temperature profile which does not satisfy the steadiness condition. 
Last, in order to self-consistently recover a steady-state solution, the temperature profile is updated by solving the following form of the energy equation (e.g. \citealt{MurrayClay2009}):
%%%%%%%%%%%%%%%%%%%%%%%%%%%%%%
\begin{equation}
    \rho v \frac{\de }{\de r}\left( \frac{k_B T}{(\gamma-1)\mu}\right) = \frac{k_BTv}{\mu}\der{\rho}{r} + Q(T),
    \label{eq:TemperatureEquation}
\end{equation}
%%%%%%%%%%%%%%%%%%%%%%%%%%%%%%
where we substitute the post-processed ionization fraction profiles in the expression of $\mu$ and $Q$, whereas $v(r)$ and $\rho(r)$ are again given by the stationary-state solutions. This enables us to solve for only one unknown, namely $T(r)$ (where the boundary condition is the same as the one adopted for the temporal evolution; see \S~\ref{sec:Setup} for details). 

We stress that the post-processing scheme described above is only carried out once at the end of each simulation. Overall, this approach yields very good agreement with the ionization and temperature profiles obtained through TPCI (S16), and does so at a fraction of the computing time. 

%%%%%%%%%%%%%%%%%%%%%%%%%%%%%%%%%%%%%%%%%%%%%%%%%%

\subsection{Two-dimensional effects}
\label{subsec:2Deffects}
Even under the assumption of parallel rays (i.e., infinite distance to the star), the geometry of the radiative transfer problem is intrinsically two-dimensional, as the irradiating stellar photons see different optical depths across different atmospheric angles from the substellar point. Moreover, the photo-heating and ionization rates ought to be averaged over the planet day-side \citep[see, e.g.,][]{Erkaev2013}. 

To simplify this, \citet{odert20} modifies the photo-heating rate by dividing the stellar flux by a factor $(1+\alpha\tau_E)$ \citep{Sekiya1980}, arguing that, in the case of HD~189733~b, the solution approximates well the averaged 2D case for $\alpha=4$.
Instead, S16 adopt the same photo-heating rate as in Equation~\ref{eq:HeatingRate1D}, and then divide the resulting, steady-state mass loss rate $\dot M=4\pi\rho v r^2 $ by a factor of $4$, in order to account for the day-side illumination and evaporation. 

A somewhat different approach--that we propose--stems from the comparison between a tidally-locked planet and a rapidly spinning one.
In the former case, that we adopt in all simulations, the day-side averaged rates will be given by Equations~\ref{eq:HeatingRate1D} and~\ref{eq:PhotoionizationRate1D}--both divided by a factor of $2$--while the mass outflow originates from one side of the planet, i.e., $\dot M=2\pi\rho v r^2 $. For a rapidly spinning planet, instead, the mass outflow originates from the entire planet, whereas the photo-heating rates will be reduced by a factor of $4$ compared to Equations~\ref{eq:HeatingRate1D} and~\ref{eq:PhotoionizationRate1D} (note, however, that photoevaporation may be negligible for planets far enough from their stars that they could/should be treated as rapidly rotating; for context, a thorough discussion of the two-dimensional effects of irradiation for tidally-locked planets compared to rapidly spinning ones can be found in \citealt{Showman2015}).

ATES users have the option to select one amongst the above-mentioned recipes. 
We defer to \S\ref{sec:Results} for a discussion of the effects of these different approaches to the simulated profiles and mass loss rates. 
%The reader is referred to \citealt{showman15} for a three-dimensional analysis of the combined effects of rotation and orbital separation applied to atmospheric circulation in gas giants. 
%-----------------------------------------------------------------------------

\section{Numerical methods}
\label{sec:NumericalMethods}

ATES is a Fortran 90/95 Godunov-type hydrodynamical code that solves the spherical Euler equations numerically, through a finite-volume discretization. ATES has been developed specifically for the study of atmospheric evaporation in exoplanets, although it could be easily modified for the purpose of simulating more general astrophysical phenomena. The code is built in a modular fashion which allows for the straightforward inclusion (and modification/addition) of different physical processes, e.g., gravity, radiation effects, and chemistry. Its main features are detailed below. 

\subsection{Spatial grid}
The spatial domain extends from $R_p$, the planetary radius, to $R_R$, the system Roche lobe radius. It is discretized into $N=500$ computational cells. Three different spatial grid types are implemented, and classified as follows:
\begin{enumerate}
    \item Uniform grid: $\Delta r_j = (R_R - R_p)/N$, which is suitable for problems that do not involve large gradients in the flow parameters close to $R_p$;
    \item Stretched grid: $\Delta r_j = k\Delta r_{j-1}, ~ k = R_R^{1/N}$, which is suitable for solutions with moderately high gradients close to $R_p$;
    \item Mixed uniform-stretched grid: 
    %%%%%%%%%%%%%%%%%%%%%%%%%%%%%%%%%%%%%%%%%
    \begin{equation}
        \Delta r_j = 
        \begin{cases}
            \begin{aligned}
                & \Delta r_{low} = 2\cdot 10^{-4}R_p && \quad \text{if}~j<50\\
                & k\Delta r_{j-1} && \quad \text{if} ~j\geq 50
            \end{aligned}
        \end{cases},
    \end{equation}
    %%%%%%%%%%%%%%%%%%%%%%%%%%%%%%%%%%%%%%%%%
    where $k$ is evaluated by solving numerically the following equation through the Newton-Raphson method:
    %%%%%%%%%%%%%%%%%%%%%%%%%%%%%%%%%%%%%%%%%
    \begin{equation}
    \dfrac{1-k^{N-50}}{1-k} = \dfrac{R_R-50\Delta r_{low}}{\Delta r_{low}}.
    \end{equation}
    %%%%%%%%%%%%%%%%%%%%%%%%%%%%%%%%%%%%%%%%%
    The last choice is suitable for large gradients close to $R_p$; this is the default choice in ATES. 
\end{enumerate}

\subsection{Temporal and spatial discretization}

ATES uses the third-order Strong Stability Preserving Runge-Kutta method \citep[SSPRK3; ][]{Gottlieb1998} for time discretization. We indicate the vectors of cell-averaged conservative variables as $\mathbf{U} = (\rho, \rho v, E)^\intercal$, while $\mathbf{L}$ and $\mathbf{S}$ indicate the spatial operators for the hyperbolic and the source terms, respectively. The time integration scheme is written as follows:
%%%%%%%%%%%%%%%%%%%%%%%%%%%%%%%%%%%%%%%%%
\begin{equation}
    \begin{dcases}
            \mathbf{U}_j^{(1)} = \mathbf{U}_j^n + \Delta t^n( \mathbf{L}^n_j + \mathbf{S}^n_j)\\
            \mathbf{U}_j^{(2)} = \frac{3}{4}\mathbf{U}_j^n + \frac{1}{4}\mathbf{U}_j^{(1)} + \frac{\Delta t^n}{4}( \mathbf{L}^{(1)}_j + \mathbf{S}^{(1)}_j) \\
            \mathbf{\tilde{U}}_j^{n+1} = \frac{1}{3}\mathbf{U}_j^n + \frac{2}{3}\mathbf{U}_j^{(2)} + \frac{\Delta t^n}{3}( \mathbf{L}^{(2)}_j + \mathbf{S}^{(2)}_j) \\
        \end{dcases}
\end{equation}
%%%%%%%%%%%%%%%%%%%%%%%%%%%%%%%%%%%%%%%%%
The convective operators are discretized according to the standard, conservative, finite-volume procedure for spherically symmetric flows. In particular, in the $j$-th cell, the components of $\mathbf{L}$ are evaluated by the following relations:
%%%%%%%%%%%%%%%%%%%%%%%%%%%%%%%%%%%%%%%%%
\begin{equation}
    \mathbf{L}_j := 
    \begin{pmatrix}
        L^\rho_j\\[5pt]
        L^{\rho v}_j\\[5pt]
        L^E_j\\
    \end{pmatrix}
    = -\frac{\mathcal{A}_{j+1/2}\mathbf{F}_{j+1/2} - \mathcal{A}_{j-1/2}\mathbf{F}_{j-1/2}}{\Delta \mathcal{V}_j},
\end{equation}
%%%%%%%%%%%%%%%%%%%%%%%%%%%%%%%%%%%%%%%%%
where $\mathbf{F} = (\rho v, \rho v^2, v(E+p))^\intercal$ is the flux vector, $\mathcal{A}_{j\pm 1/2} = 4\pi r_{j\pm 1/2}^2$ is the area of the $(j\pm 1/2)$-th cell interface, $\Delta r_j = r_{j+1/2}-r_{j-1/2}$ and $\Delta \mathcal{V}_j = 4/3\pi (r_{j+1/2}^3 - r_{j-1/2}^3)$ are the width and the volume of the $j$-th cell, respectively. The value of the numerical flux at the interfaces is evaluated by a suitable Riemann solver:
%%%%%%%%%%%%%%%%%%%%%%%%%%%%%%%%%%%%%%%%%
\begin{equation}
    \mathbf{F}_{j\pm 1/2} := 
    \begin{pmatrix}
        F^\rho_{j\pm 1/2}\\[5pt]
        F^{\rho v}_{j\pm 1/2}\\[5pt]
        F^E_{j\pm 1/2}
    \end{pmatrix}
    \approx \mathcal{RP}(\mathbf{U}_{j\pm 1/2}^L,\mathbf{U}_{j\pm 1/2}^R),
\end{equation} 
%%%%%%%%%%%%%%%%%%%%%%%%%%%%%%%%%%%%%%%%%
where $\mathbf{U}^{L,R}_{j\pm 1/2}$ are the left and right reconstructed states at the interfaces (see \S~\ref{subsec:Recontruction}).

In order to enforce energy conservation, the third component of the flux vector is evaluated as:
%%%%%%%%%%%%%%%%%%%%%%%%%%%%%%%%%%%%%%%%%
\begin{equation}
F^E_{j\pm 1/2} = [v(E+p)]_{j\pm 1/2} + F^\rho_{j\pm 1/2}\Phi_{j\pm 1/2},
\end{equation}
%%%%%%%%%%%%%%%%%%%%%%%%%%%%%%%%%%%%%%%%%
with $\Phi_{j\pm 1/2} \equiv \Phi(r_{j\pm 1/2})$. The pressure gradient in the momentum equation is included in the source vector, which is evaluated as follows:  
%%%%%%%%%%%%%%%%%%%%%%%%%%%%%%%%%%%%%%%%%
\begin{equation}
    \mathbf{S}_j = 
    \begin{pmatrix}
        0\\[5pt]
        -\dfrac{\rho_{j + 1/2} + \rho_{j-1/2}}{2}\dfrac{\Phi_{j+1/2} - \Phi_{j-1/2}}{\Delta r_j}\\[5pt]
        \Phi_jL^\rho_j
    \end{pmatrix}-
    \begin{pmatrix}
        0\\[5pt]
        \dfrac{p_{j+1/2} - p_{j-1/2}}{\Delta r_j}\\[5pt]
        0
    \end{pmatrix},
\end{equation}
%%%%%%%%%%%%%%%%%%%%%%%%%%%%%%%%%%%%%%%%%
where $\rho_{j\pm 1/2}$ and $p_{j\pm 1/2}$ are evaluated at the cell interfaces.\\

Radiative contributions to the conservation of energy are integrated in time by performing an explicit Euler step after the hydrodynamical evolution. At each step, the total energy density is updated, i.e.,
%%%%%%%%%%%%%%%%%%%%%%%%%%%%%%%%%%%%%%%%%
\begin{equation}
    E^{n+1} = \tilde{E}^{n+1} + \Delta t^nQ^n.
\end{equation}
%%%%%%%%%%%%%%%%%%%%%%%%%%%%%%%%%%%%%%%%%
where $Q$ is the source function defined in \S~\ref{subsec:RadiativeMechanisms}. The time-step is chosen according to the Courant-Friedrichs-Lewy (CFL) condition \citep{Toro2009}:
%%%%%%%%%%%%%%%%%%%%%%%%%%%%%%%%%%%%%%
\begin{equation}
    \Delta t^n = C_{\text{CFL}} \min\limits_j\left(\frac{\Delta r_j}{\vert v\vert_{j}^n+c_{s,j}^n}\right).
\end{equation}
%%%%%%%%%%%%%%%%%%%%%%%%%%%%%%%%%%%%%%
Here, $c_s$ is the local sound speed, and $C_{\text{CFL}}=0.6$.

\subsection{Approximate Riemann solvers}
\label{subsec:ApproxRiemann}
The flux at the cell interface is computed by means of approximate Riemann solvers. Hereafter, $\mathbf{U}_{L}$ ($\mathbf{U}_{R}$) will denote the states on the left (right) with respect to the  $(j+1/2)$-th interface. 
%%%%%%%%%%%%%%%%%%%%%%%%%%%%%%%%%%%%%%%
\begin{enumerate}

\item The Harten-Lax-Van Leer with restored contact wave flux \citep[HLLC; ][]{Toro1994}, based on a three-wave model for the structure of the exact solution of the Riemann problem. The HLLC flux is given by
%%%%%%%%%%%%%%%%%%%%%%%%%%%%%%%%%%%%%%%
\begin{equation}
    \mathbf{F}_{j+1/2} = 
        \begin{cases}
            \begin{aligned}
                     & \mathbf{F}_L ~ && 0\leq S_L\\
                     & \mathbf{F}_{*L} ~ && S_L\leq 0\leq S_*\\
                     & \mathbf{F}_{*R} ~ && S_*\leq 0\leq S_R\\
                     & \mathbf{F}_R ~ && 0\geq S_R\\
            \end{aligned}
        \end{cases},
\end{equation}
%%%%%%%%%%%%%%%%%%%%%%%%%%%%%%%%%%%%%%%
where $\mathbf{F}_{*K} = \mathbf{F}_K + S_K(\mathbf{U}_{*K} - \mathbf{U}_K)$, with $K=L,R$, and
\begin{equation}
    \mathbf{U}_{*K} = \rho_K\left( \frac{S_K-v_K}{S_K-S_*}\right)
    \begin{pmatrix}
        1\\[5pt]
        S_*\\[5pt]
        \dfrac{E_K}{\rho_K} + (S_*-v_K)\left[ S_* + \dfrac{p_K}{\rho_K(S_K-v_K)}\right]\\
    \end{pmatrix}.
\end{equation}
%%%%%%%%%%%%%%%%%%%%%%%%%%%%%%%%%%%%%%%
The three characteristic speeds are chosen from \citet{Batten1997};
%%%%%%%%%%%%%%%%%%%%%%%%%%%%%%%%%%%%%%%
\begin{equation}
    \begin{aligned}
        S_L & = \min(0,v_L-c_{sL},v_R-c_{s R})\\
        S_R & = \min(0,v_L+c_{sL},v_R+c_{s R})\\
        S_* & = \frac{p_R-p_L+\rho_Lv_L(S_L-v_L)-\rho_Rv_R(S_R-v_R)}{\rho_L(S_L-v_L)-\rho_R(S_R-v_R)}.
    \end{aligned}
\end{equation}
This scheme is able to handle solutions containing both smooth regions as well as discontinuities which may arise during the temporal evolution; this is the default choice in ATES.

\item The Roe solver \citep{Roe1981}, which is based on the exact solution of a local linearized Riemann problem at the interface. In this case, the flux is given by:
%%%%%%%%%%%%%%%%%%%%%%%%%%%%%%%%%%%%%%%
\begin{equation}
    \mathbf{F}_{j+1/2} = \frac{1}{2}\left(\mathbf{F}_L + \mathbf{F}_R\right) - \frac{1}{2}\sum\limits_{i=1}^3{\tilde{\alpha}_i\vert\tilde{\lambda}_i\vert \mathbf{\tilde{K}}^{(i)}}, 
\end{equation}
%%%%%%%%%%%%%%%%%%%%%%%%%%%%%%%%%%%%%%%
where the wave strengths $\tilde{\alpha}_i$, the eigenvalues $\tilde{\lambda}_i$ and the eigenvectors $\mathbf{\tilde{K}}^{(i)}$ of the Jacobian are evaluated on a reference state defined from averages \citep[e.g.][]{Toro2009}:
\begin{equation}
    \begin{pmatrix}
        \tilde{\rho}\\[5pt]
        \tilde{v}\\[5pt]
        \tilde{H}
    \end{pmatrix}=
    \begin{pmatrix}
        \sqrt{\rho_L\rho_R}\\[5pt]
        \dfrac{\sqrt{\rho_L}v_L + \sqrt{\rho_R}v_R}{\sqrt{\rho_L} + \sqrt{\rho_R} }\\[10pt]
        \dfrac{\sqrt{\rho_L}H_L + \sqrt{\rho_R}H_R}{\sqrt{\rho_L} + \sqrt{\rho_R} }
    \end{pmatrix}.
\end{equation}
%%%%%%%%%%%%%%%%%%%%%%%%%%%%%%%%%%%%%%%
Here $H = (E+p)/\rho$ is the specific entalpy of the gas. The code includes also the Harten-Hyman entropy. This option is suitable for those cases where possible discontinuities ought to be resolved with high accuracy.

\item The Local Lax-Friedrichs flux \citep[LLF; ][]{Rusanov1962}, in which the numerical flux is evaluated as
        %%%%%%%%%%%%%%%%%%%%%%%%%%%%%%%%%%%%%%%
        \begin{equation}
            \mathbf{F}_{j+1/2} = \frac{\mathbf{F}_L + \mathbf{F}_R}{2} - \alpha_{j+1/2}\left( \mathbf{U}_R - \mathbf{U}_L\right),
        \end{equation}
        %%%%%%%%%%%%%%%%%%%%%%%%%%%%%%%%%%%%%%%
        where $\alpha_{j+1/2} = \max(\vert v_L\vert + c_{sL}, \vert v_R \vert + c_{sR})$ is the maximum local eigenvalue of the Jacobian of the Euler's equations (Equation~\ref{eq:EulerEquations}).
The LLF solver is the least computationally expensive option; however, it is also the most diffusive, and thus should be used only in conjunction with a high resolution reconstruction scheme (see \S~\ref{subsec:Recontruction}). 

\end{enumerate}

\subsection{Reconstruction}
\label{subsec:Recontruction}

The accuracy of the reconstruction of the left and right states at a given interface determines the overall spatial accuracy of the numerical scheme. Within ATES, the reconstruction is carried out in primitive variables $\mathbf{W} = (\rho, v, p)^\intercal$. Two reconstruction schemes are available to the user:
%%%%%%%%%%%%%%%%%%%%%%%%%%%%%%
\begin{enumerate}
    \item A second-order Piecewise Linear reconstruction Method \citep[PLM; see, e.g., ][]{Leveque2002}. The states at the interfaces of the $j-$th cell are reconstructed through a linear interpolation on the stencils $\{j-1,j\}$ and $\{j,j+1\}$:
    %%%%%%%%%%%%%%%%%%%%%%%%%%%%%%
    \begin{equation}
        \begin{cases}
            \begin{aligned}
                \mathbf{W}_{R,j-1/2} & = \mathbf{W}_j - \frac{1}{2}\sigma \Delta r_{j-1/2}\\
                \mathbf{W}_{L,j+1/2} & = \mathbf{W}_j +\frac{1}{2}\sigma \Delta r_{j+1/2}
            \end{aligned},
        \end{cases}
    \end{equation}
    %%%%%%%%%%%%%%%%%%%%%%%%%%%%%%
    where $\sigma$ is a limited slope evaluated through the generalized \textsc{MinMod} limiter \citep{Kurganov2000} with $\theta = 2$,
    %%%%%%%%%%%%%%%%%%%%%%%%%%%%%%
    \begin{equation}
        \sigma = \text{\textsc{MinMod}}\left(
        \theta\dfrac{\Delta \mathbf{W}_{j+1/2}}{\Delta r_{j+1/2}} ,
        \theta\dfrac{\Delta \mathbf{W}_{j-1/2}}{\Delta r_{j-1/2}},
        \dfrac{\Delta \mathbf{W}_{j+1/2} + \Delta \mathbf{W}_{j-1/2}}{\Delta r_{j+1/2} + \Delta r_{j-1/2}}  
         \right).
    \end{equation}
    %%%%%%%%%%%%%%%%%%%%%%%%%%%%%%
    The $\text{\textsc{MinMod}}$ function is defined as 
    \begin{equation}
        \text{\textsc{MinMod}}(x_1,x_2,...) = 
        \begin{cases}
            \begin{aligned}
                      \min\limits_j\{x_j\} \quad & \text{if} ~ x_j>0~\forall j\\
                      \max\limits_j\{x_j\} \quad & \text{if} ~ x_j<0~\forall j\\
                      0 \quad & \text{otherwise}.
            \end{aligned}
        \end{cases}
    \end{equation}
    %%%%%%%%%%%%%%%%%%%%%%%%%%%%%%
    
    \item A third order, Energy Stable Weighted Essentially Non-Oscillatory scheme \citep[ESWENO3; ][]{Yamaleev2009,Mignone2011}. This method employs a weighted convex combination of second-order interpolants to produce a third-order parabolic reconstruction of the left and right states. Here, we follow the compact formulation of \citet{Mignone2011}. In the $j-$th cell, we calculate the reconstructed states as follows:
    %%%%%%%%%%%%%%%%%%%%%%%%%%%%%%
    \begin{equation}
        \begin{cases}
            \begin{aligned}
                \mathbf{W}_{L,j+1/2} & = \mathbf{W}_j + \frac{1}{2}\dfrac{2a_+\Delta \mathbf{W}_{j+1/2} +a_-\Delta \mathbf{W}_{j-1/2}}{2a_+ + a_-} \\[5pt]
                \mathbf{W}_{R,j-1/2} & = \mathbf{W}_j - \frac{1}{2}\dfrac{a_+\Delta \mathbf{W}_{j+1/2} +2a_-\Delta \mathbf{W}_{j-1/2}}{a_+ + 2a_-} \\ 
            \end{aligned}
        \end{cases}.
    \end{equation}
    %%%%%%%%%%%%%%%%%%%%%%%%%%%%%%
    The coefficients $a_{\pm}$ are the weights proposed by \citet{Yamaleev2009}:
    %%%%%%%%%%%%%%%%%%%%%%%%%%%%%%
    \begin{equation}
        a_\pm = 1 + \frac{\left( \Delta \mathbf{W}_{j+1/2} - \Delta \mathbf{W}_{j-1/2}\right)^2}{\left(\Delta r_j\right)^2 + \left(\Delta \mathbf{W}_{j\pm 1/2}\right)^2}.
    \end{equation}
    %%%%%%%%%%%%%%%%%%%%%%%%%%%%%%
\end{enumerate}

Both methods are used in turn; the PLM reconstruction method is be employed first, starting from the initial conditions described below (\S~\ref{sec:Setup}). Once fractional variations in the mass flux reach a reference value of $\Delta$\mdot/\mdot $\lesssim 0.5-1$, the simulation is interrupted. The output profiles are specified as initial conditions for the second part of the run, which employs the ESWENO3 reconstruction method. The simulation is stopped when the mass outflow reaches steady-state, with tolerance set to $\Delta$\mdot/\mdot$<0.001$. 
%%%%%%%%%%%%%%%%%%%%%%%%%%%%%%%%%%%
\subsection{Ionization equilibrium}
At each time-step, the simplified photoionization equilibrium equations (i.e., eq.~\ref{eq:RadiativeEquilibriumSystem1} with the l.h.s. advection terms set equal to $0$) are solved through a modification of the Powell dogleg method \citep{Powell1970}, available through the MINPACK library\footnote{\url{https://www.netlib.org/minpack/}}. The differential  Equations~\ref{eq:RadiativeEquilibriumSystem2} and Equation~\ref{eq:TemperatureEquation} are solved by means of a standard, implicit Euler method, under the assumptions discussed in \S~\ref{subsec:RadiativeMechanisms}.
%--------------------------------------------------------

\section{Code validation}
\label{sec:Validation}

Here we present two standard problems that are routinely used to validate the performance of radiation hydrodynamics numerical schemes. Specifically, we test the pure hydrodynamical discretization and the photoionization equilibrium modules separately, by modeling the classical Sedov blast wave and rarefied ionization fronts problems. 

\subsection{Sedov blast wave}
In this classical test, a certain amount of energy is suddenly released in a small region of a low-density gas, initially assumed at rest. \citet{Sedov1959} derived the homonymous, self-similar solution of the Euler equations for this problem, which describes the formation and propagation of a blast wave. This test is especially useful for validating ATES' ability to deal with strong discontinuities in spherical coordinates. On a uniform grid of $500$ cells extending in the (dimensionless) space range $[0,1/2]$, we start from a (dimensionless) initial constant density $\rho= 1$ and pressure $p=10^{-5}$. A total energy $E=1$ is deposited into the first computational cell at $r=0$. Euler equations are then solved for an elapsed time $t = 0.05$, adopting reflective boundary conditions at $r=0$ and free-flow conditions at the upper boundary of the domain.
\begin{figure}
    \centering
    \includegraphics[width=0.9 \columnwidth]{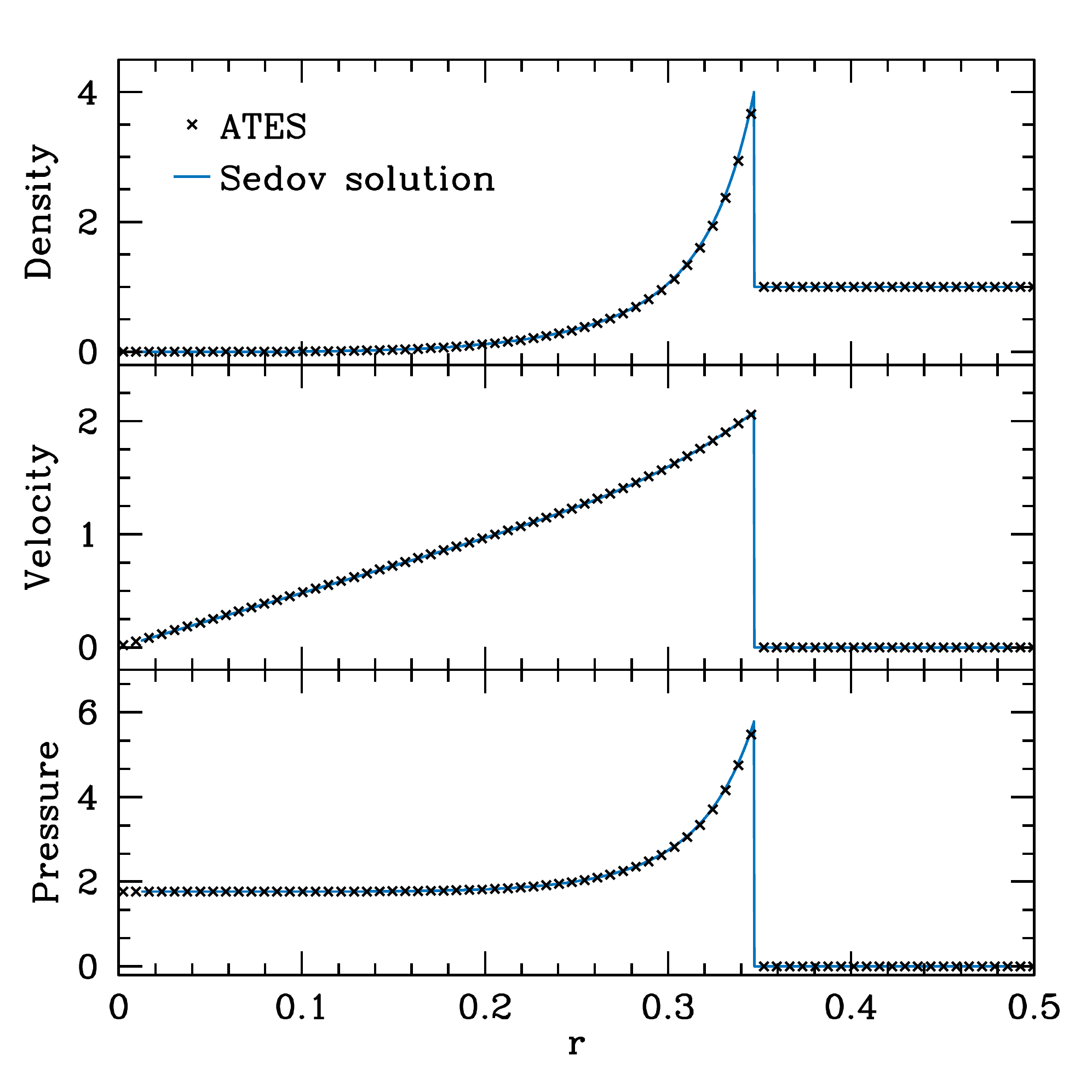}
    \caption{Sedov blast wave solution at $t=0.05$; the solid blue lines correspond to the exact solutions of the Euler equations, while the black crosses are the numerical values obtained by ATES employing the ESWENO3 reconstruction method and the HLLC approximate Riemann solver. }
    \label{fig:SedovTest}
\end{figure}
The resulting density, velocity and pressure of the blast wave, shown as black crosses in Figure \ref{fig:SedovTest}, are all in excellent agreement with the exact solutions, shown as blue lines.   

%-----------------
%%%%%%%%%%%%%%%%%%%%%%%%%%%%%%%%%%%%
\begin{figure}
    \centering
    \includegraphics[width= \columnwidth]{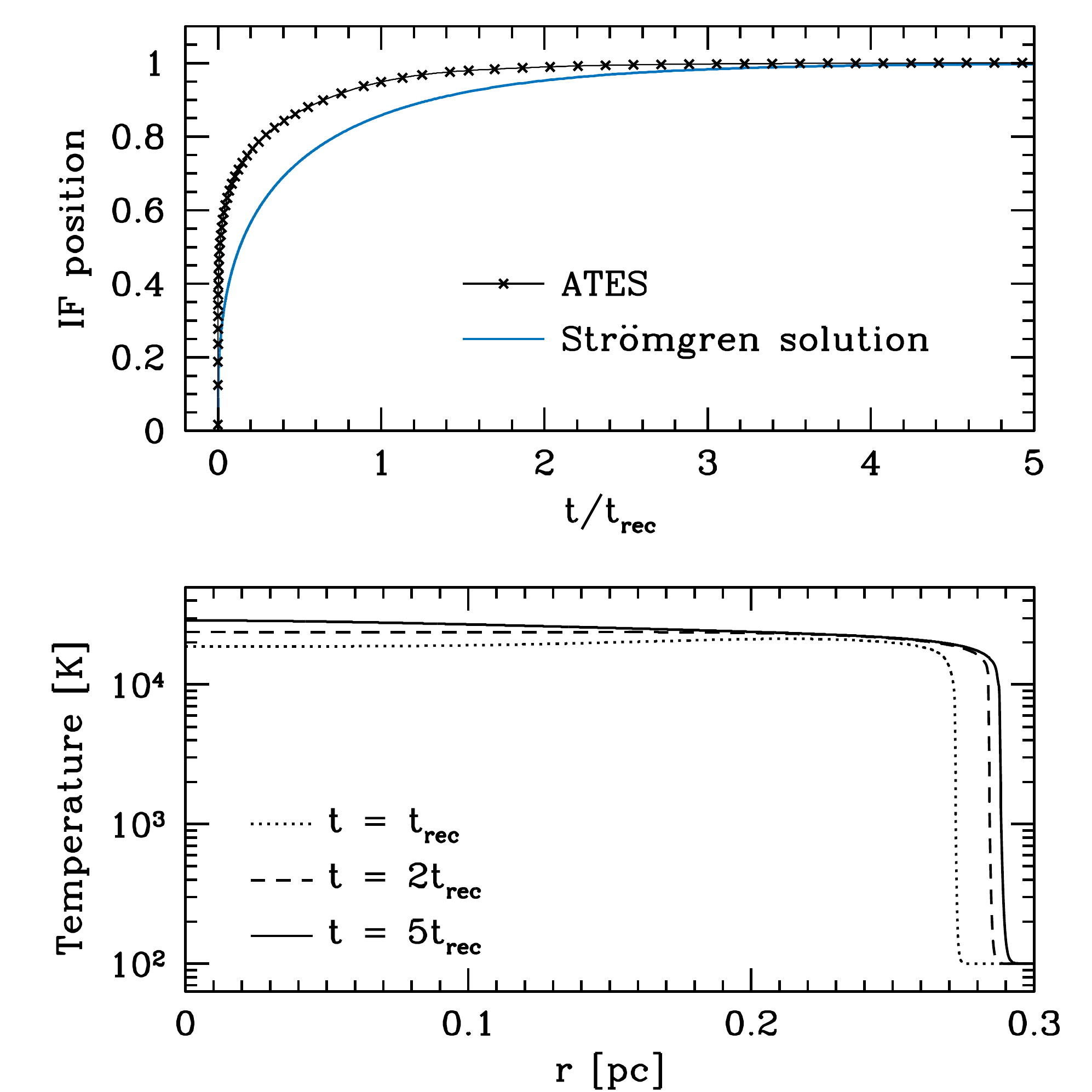}
    \caption{Top panel: temporal evolution of the ionization front. The evolution of the ionization front position (in units of $R_S$; see Equation~\ref{eq:Stromradius}) computed by ATES is represented by the black crosses, and compared to the analytical solution (see Equation~\ref{eq:StromgrRIFtime}), shown by the solid blue line. Bottom panel: the ionization front temperature profiles are shown at different times, as a function of the distance from the central star.}
    %The HLLC Riemann solver and the ESWENO3 reconstruction method are used for this test.}
    \label{fig:StromgrenTest}
\end{figure}
%%%%%%%%%%%%%%%%%%%%%%%%%%%%%%%%%%
\subsection{Rarefied ionization front}
As a second test, we simulate the formation of an \ion{H}{II} region around a central ionizing source. This is meant to validate the implementation of the numerical solver for the photoionization equilibrium equations.

A hot (O-type) star, with a $T_\star = 50\, 000$~K blackbody spectrum, is placed at the origin of the reference system. A neutral, homogeneous hydrogen nebula with constant density $n_\Hy = 10^4$~cm$^{-3}$ fills the domain, which extends up to $0.3$~pc. The nebula is initially at rest, at a temperature of $T=100$~K. A strong shock wave develops as soon as the central source is switched on. The position of the R-type Ionization Front (IF) follows the well-known result of  \citet{Stromgren1939}:
%%%%%%%%%%%%%%%%%%%%%%%%%%%%%%%%%%%%%%
\begin{equation}
    R_{\text{IF}}(t) = R_S\left( 1 - e^{-t/t_{\text{rec}}}\right)^{1/3},
    \label{eq:StromgrRIFtime}
\end{equation}
%%%%%%%%%%%%%%%%%%%%%%%%%%%%%%%%%%%%%%
where $t_{\text{rec}} = (\alpha_{\text{rec},\HII} n_e)^{-1}$ is the case-B recombination time for $\HII$, and the so-called Str\"{o}mgren radius for an isothermal sphere is found by balancing the total number of recombinations and photoionizations:
%%%%%%%%%%%%%%%%%%%%%%%%%%%%%%%%%%%%%%
\begin{equation}
    R_S = \left(\frac{3\dot N}{4\pi n_\HI^2\alpha_{\text{rec},\HII}(T)}\right)^{1/3}.
    \label{eq:Stromradius}
\end{equation}
%%%%%%%%%%%%%%%%%%%%%%%%%%%%%%%%%%%%%%
Here, $\dot N$ is the number of ionizing photons emitted per second by the star. For a stellar radius of $10$ $R_\odot$, we find $\dot N\simeq 3.8\times 10^{49}$ s$^{-1}$. Assuming a gas temperature of $T=2.2\times 10^4$~K, we find $\alpha_{\text{rec},\HII} \simeq 1.31 \times 10^{-13}$~cm$^3$s$^{-1}$, while the inferred Str\"{o}mgren radius is $R_S \simeq 0.29$~pc. 

We adopt a radial grid of $N = 1000$ points, with zero-gradient boundary conditions on all variables. The solution is advanced in time up to $5\,t_{\text{rec}}$. The post-processing routine described in \S~\ref{sec:ModelDescription} (to account for advection) is not used here. The time-step is limited by the minimum recombination time value in the domain, i.e., $\Delta t_{\text{rec}}^n = [\max(n_{e}^n \alpha_{\text{rec},\HII}(T^n))]^{-1}$, and by ensuring the positivity of the internal energy, i.e., $\Delta t^n_{\text{rad}} = 0.9 \min{( E^n/\vert Q^n \vert)}$. We thus choose $\Delta t^n = \min(\Delta t_{\text{rec}}^n,\Delta t^n_{\text{rad}})$.

As shown in the top panel of Figure~\ref{fig:StromgrenTest}, the numerical IF (black crosses) propagates at a faster pace compared to the analytical solution (blue line) during the early stages of the evolution. During this phase, the temperature of the \HII~ region is significantly lower than the final equilibrium temperature, and the resulting recombination timescale is longer than the radiative timescale, i.e., ionization and recombination are out of equilibrium. As ATES computes the ionization fractions under an assumed equilibrium, the IF velocity is necessarily overestimated for $t\ll t_{\text{rec}}$, when in fact equilibrium is actually not settled. It is only at later times, when the temperature reaches $\sim 10^4$~K, as indicated by the solid line in the lower panel of Figure \ref{fig:StromgrenTest}, that the recombination and radiative timescales become comparable, and the asymptotic value of the IF position evaluated by ATES matches the Str\"omgren radius.  

%-------------------------------------------------------------

%%%%%%%%%%%%%%%%%%%%%%%%%%%%%%%%%%%%%%%%%%%%%%%%%%
\begin{table*}
\renewcommand{\arraystretch}{1.2}
\small
\centering
\caption{Planetary systems considered in this work.}
\label{tab:planets}
\makebox[\textwidth][l]{
\begin{tabular}{
	P{1.9cm} % Planet name
	P{0.7cm} % Rp
	P{0.7cm} % Mp
	P{0.5cm} % T0 P{0.5cm}
	P{0.7cm} % a P{0.7cm}
	P{1.0cm} % Mdot
	P{1pt}   % empty
	P{1.0cm} % M_star
	P{1.2cm} % L_X
	P{1.2cm} % L_EUV
	}
	\hline
	\hline
	% Header
	%	 &  \multicolumn{4}{c}{\textbf{Planet}} & & \multicolumn{3}{c}{\textbf{Host star}} \\
		 %   \cline{2-6} \cline{8-10}
		
	% Contents
	%\textbf{Planet}   																						& % 
	&
	\boldmath$R_p$ \newline $\mathbf{[R_J^*]}$ 																& %
	\boldmath$M_p$ \newline  $\mathbf{[M_J]}$ 																& %
	\boldmath$T_p$ \newline $\mathbf{[K]}$																	& % 
	\boldmath$a$ \newline $\mathbf{[AU]}$																	& %
	\boldmath$\log\dot M$\newline \textbf{[g s$^\mathbf{-1}$]}                                        & &
	\boldmath$M_\star$ \newline $\mathbf{[M_\odot]}$													& %
	\boldmath$\log L_\mathbf{X}$ \newline \textbf{[erg s$\mathbf{^{-1}}$]}   							& %
	\boldmath$\log L_\mathbf{EUV}$ \newline \textbf{[erg s$\mathbf{^{-1}}$]} 							\\% #14
	& (1) & (2) & (3) & (4) & (5) & & (6) & (7) & (8) \\
	 \cline{2-10}
	% Parameters
    % planet             Rp        Mp    T0       a    Mdot    Ms       Lx    LEUV          
    GJ 1214	b	      & 0.24 & 0.020 &	550	& 0.014 & {\bf 9.83}\,(9.68)   & & 0.15  & 25.91  & 26.61  \\
    HD 97658 b		  & 0.21 & 0.025 &  750	& 0.080 & {\bf 9.58}\,(9.47)   & & 0.850 & 27.22  & 28.19  \\
    55 Cnc e		  & 0.19 & 0.026 & 1950 & 0.015 & {\bf 10.33}\,(10.14) & & 1.015 & 26.65  & 27.66  \\
    GJ 436 b	 	  & 0.38 & 0.073 &	650	& 0.029 & {\bf 9.65}\,(9.65)   & & 0.452 & 25.96  & 27.14  \\
    HAT-P-11 b        & 0.42 & 0.083 &	850	& 0.053 & {\bf 10.36}\,(10.29) & & 0.809 & 27.55  & 28.33  \\
    WASP-80	b	      & 0.95 & 0.55  &	800	& 0.034 & {\bf 10.74}\,(10.55) & & 0.580 & 27.85  & 28.46  \\
    HD 209458 b	      & 1.4  & 0.69  & 1320	& 0.047 & {\bf 10.54}\,(10.27) & & 1.148 & <26.40 & <27.84 \\
    HD 189733 b	      & 1.1  & 1.1   & 1200 & 0.031 & {\bf 9.97}\,(9.61)   & & 0.800 & 28.18  & 28.61  \\
    WASP-77 A b	      & 1.2  & 1.8   & 1650 & 0.024 & {\bf 9.07}\,(8.79)   & & 1.002 & 28.13  & 28.59  \\
    WASP-43	b	      & 0.93 & 1.8   & 1350 & 0.014 & {\bf 8.50}\,(8.04)   & & 0.717 & 27.88  & 28.48  \\
    WASP-12 b         & 1.8  & 1.4   & 2900 & 0.023 & {\bf 11.87}\,(11.60) & & 1.434 & <27.58 & <28.35 \\
    CoRoT-2 b         & 1.5  & 3.3   & 1550 & 0.028 & {\bf 7.69}\,(7.63)   & & 0.97  & 29.32  & 29.13  \\
    GJ 3470 b         & 0.37 & 0.044 & 650  & 0.036 & {\bf 10.76}\,(10.66) & & 0.51  & 27.63  & 28.37  \\
    HD 149026 b       & 0.65 & 0.36  & 1440 & 0.043 & {\bf 10.79}\,(10.43) & & 1.3   & 28.60  & 28.80  \\[2pt]
	\hline
	\hline
\end{tabular}
}
\tablefoot{Explanation of the columns: planet's radius (1), mass (2), zero-albedo equilibrium temperature (3), average orbital distance$^{(1)}$ (4), ATES (in bold) vs. TCPI (in brackets) mass outflow rate (5), host star mass (6), X-ray (7), and EUV luminosity (8).
}
\tablebib{The planets considered in this work, along with their parameters, are drawn from S16, and references therein. Starting from a list of 21 targets (see table 1 in S16), we limit our list to those which are estimated to have out-flowing (rather than stable) atmospheres, i.e. targets 1--14 in table 3 of S16. The mass outflow rates obtained by ATES are shown in boldface, next to the TPCI values by S16, in brackets (see Figure~\ref{fig:mdot} for a visual comparison). $^*$Refers to the volumetric mean radius $R_J=6.99\times 10^9$ cm.}
\end{table*}
%%%%%%%%%%%%%%%%%%%%%%%%%%%%%%%%%%%%%%%%%%%%%%%%%%

%
%
\section{Simulation setup}
\label{sec:Setup}

In order to run ATES for a specific exoplanet, Euler's equations (Equation~\ref{eq:EulerEquations}) are recast in dimensionless form, with the planet radius $R_P$ serving as unit length and $\sqrt{k_B T_p/m_\Hy}$ as unit velocity, $T_p$ being the gas temperature at the planet radius (the so-called planet equilibrium temperature with zero albedo).

The planet atmosphere is then initialized as a fully neutral, isothermal sphere composed by \Hy and \He (with number ratio set to the cosmological value of $1/12$ in the cases presented here), with a liner initial velocity profile: $v(t=0) = (r-1)/2$.

The atmospheric density and temperature at the planet radius are chosen according to realistic physical conditions. The total number density at planet radius is set to $10^{14}$ cm$^{-3}$, the same value adopted by S16. The reader is referred to their Section 3.6 for a detailed discussion of the effects the chosen value has on the stationary solutions. The equilibrium temperature depends on the specific planet. For the runs describe in next \S~\ref{sec:Results}, again we adopt the same values as S16 (see their table 1 for references). The value of the velocity in the ghost cells at the lower boundary is obtained through a zero-th order extrapolation from the first computational cell in the case of inflow characteristics, while it is set to zero otherwise. At the upper boundary (i.e. at the Roche lobe radius), a zero-gradient condition is applied to all quantities if the PLM reconstruction method is employed; in the case of the ESWENO3 reconstruction, the ghost cells are filled with linearly extrapolated values from the outermost cell of the domain. 

The stellar spectrum is modelled as a piece-wise power-law with the specific flux $\propto E^{-1}$, both in the EUV band ($[13.6,124]$ eV), and the X-ray band ($[0.124,12.4]$ keV); the spectrum in each band is normalized to the EUV and X-ray luminosities reported by S16 for each planet. The validity of this approximation is discussed in \S~\ref{sec:Results}.

All the simulations presented in this work were performed employing: the mixed uniform-stretched grid, HLLC approximate Riemann solver and the PLM--followed by ESWENO reconstruction scheme (see \S~\ref{sec:NumericalMethods} above for details).
A typical simulation runs for few tens of minutes on a 2.7 GHz quad-core CPU, with some cases taking as little as a few minutes, up to a few hours for the most time-consuming cases. 
By comparison, S16 report that ``The computational effort of the presented simulations corresponds approximately to 300,000 h on a standard 1 GHz CPU''; this refers to all of the 18 planets listed in their table 1.

\begin{figure*}
    \centering
    \includegraphics[width = 1.5\columnwidth]{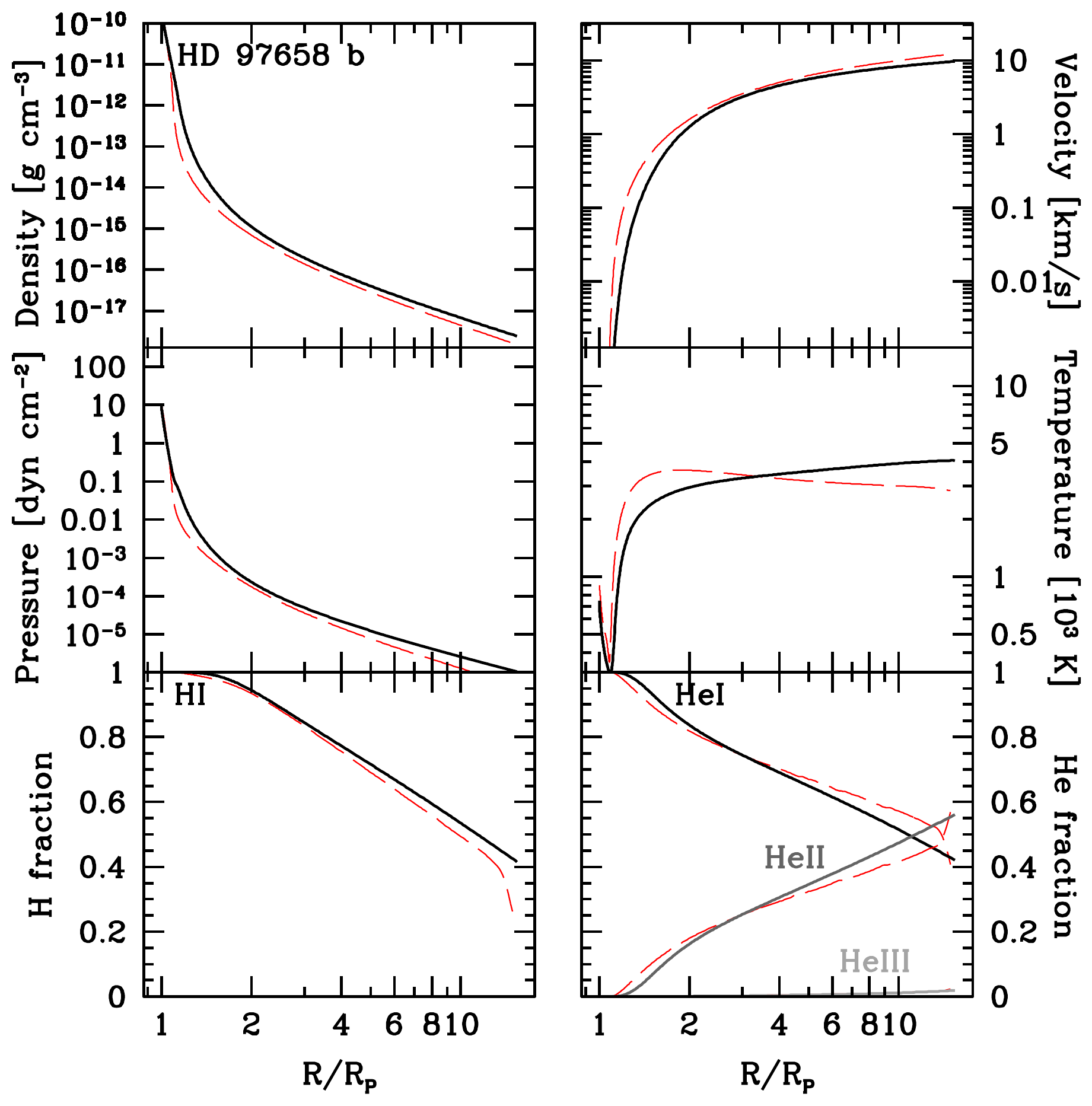}
    %\vspace{-3cm}
	\caption{Simulated density, velocity, pressure, temperature, neutral hydrogen (\HI), neutral helium (\HeI), single-ionized helium (\HeII), and double ionized helium (\HeIII) fractions, for the case of HD~97658~b. The thick solid black lines trace the profiles as calculated by ATES, to be compared to those obtained by TPCI (S16), shown as dashed red lines. Note that the \HeIII~ curves in the bottom right panel practically overlap. }
\label{fig:hd-six}
\end{figure*}
%%%%%%%%%%%%%%%%%%%%%%%%%%%%%%%%%%%%%%%%%%%%%%%%%%
\begin{figure*}
    \centering
    \includegraphics[width = 1.55\columnwidth]{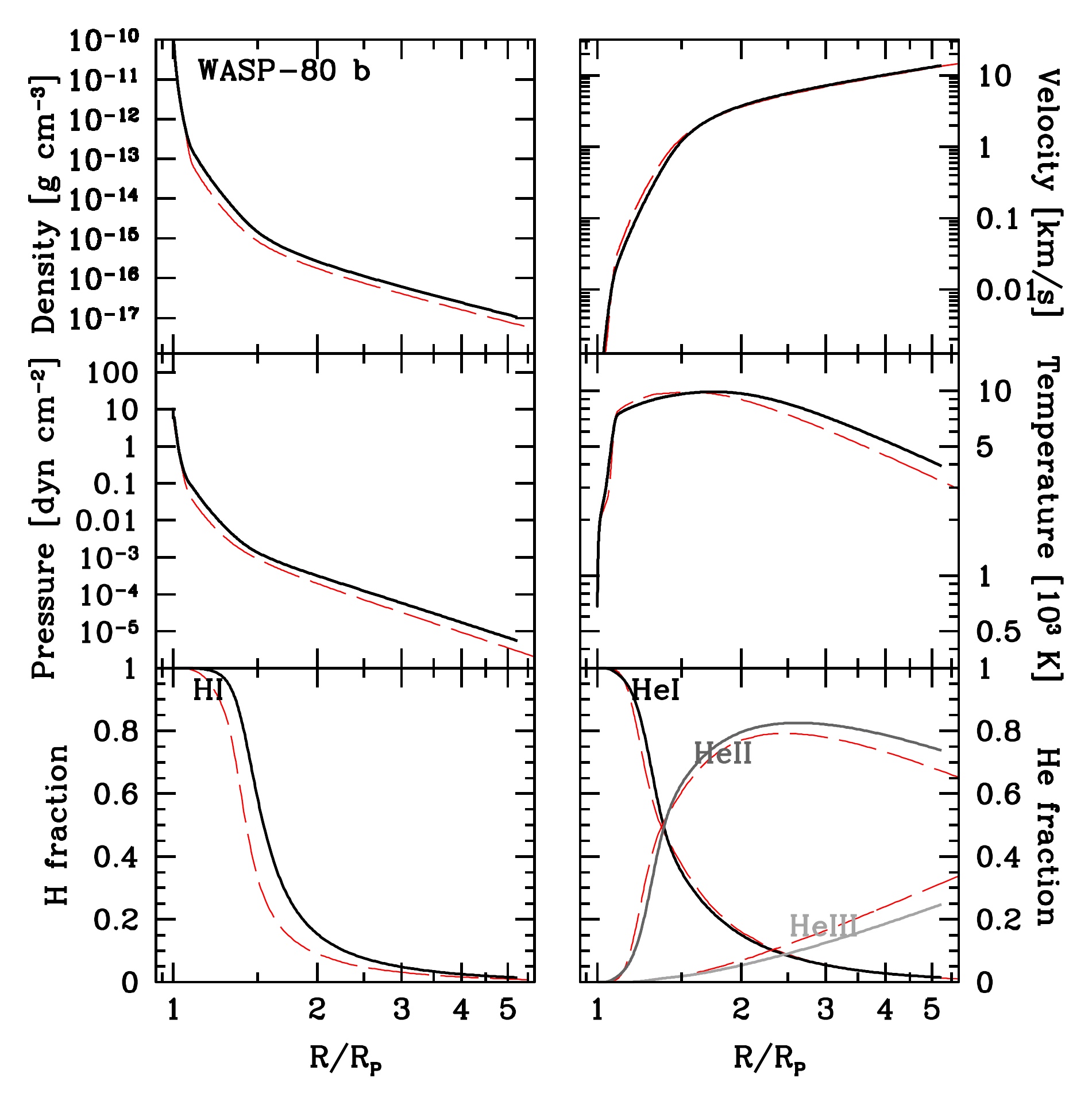}
    %\vspace{-3cm}
	\caption{Same as for Figure \ref{fig:hd-six}, but for WASP-80 b.
}
\label{fig:wasp80-six}
\end{figure*}
%%%%%%%%%%%%%%%%%%%%%%%%%%%%%%%%%%%%%%%%%%%%%%%%%%
\begin{figure*}
    \centering
    \includegraphics[width =1.5\columnwidth]{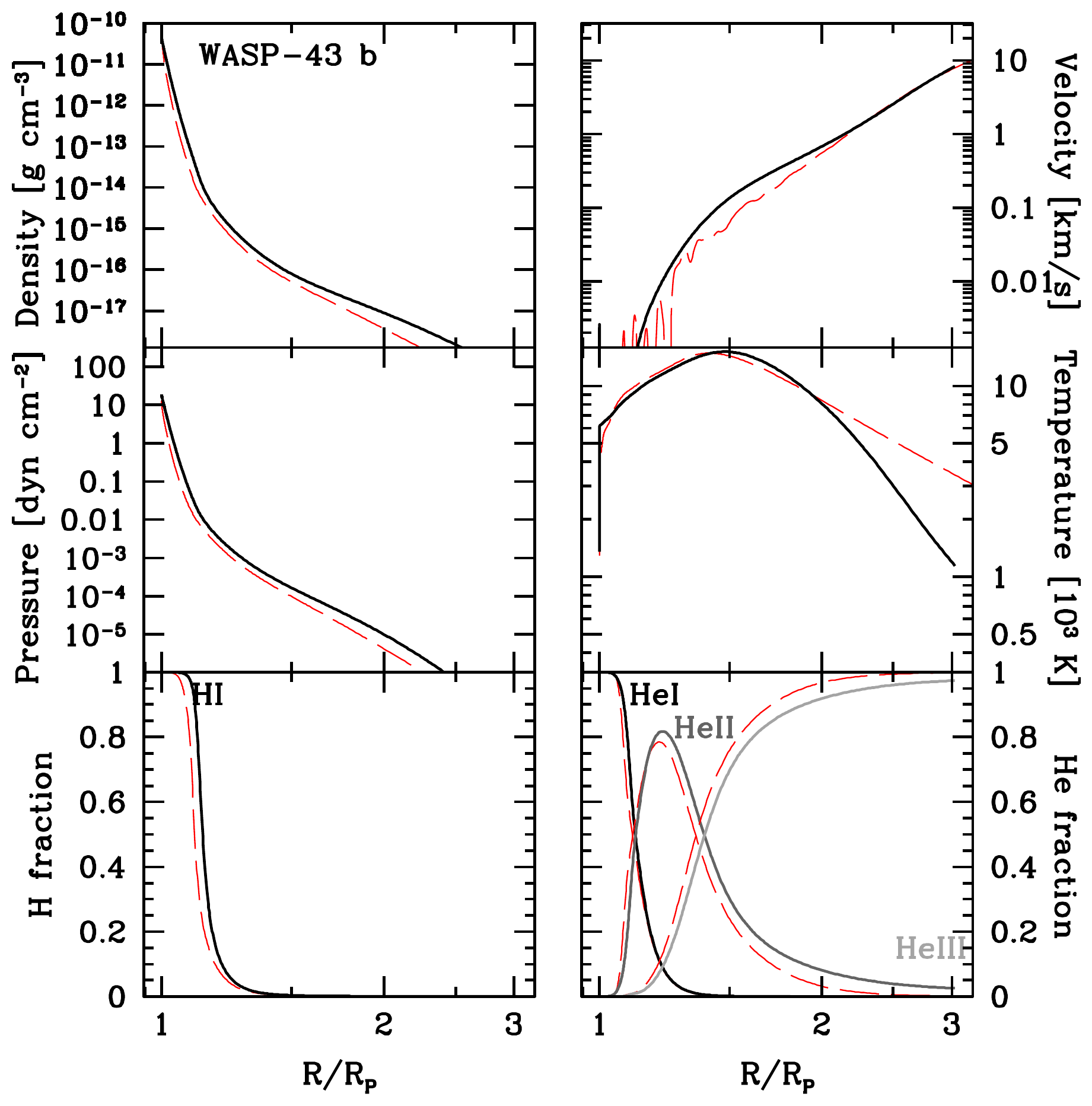}
    %\vspace{-3cm}
	\caption{Same as for Figure \ref{fig:hd-six}, but for WASP-43~b.}
\label{fig:wasp43-six}
\end{figure*}
%%%%%%%%%%%%%%%%%%%%%%%%%%%%%%%%%%%%%%%%%%%%%%%%%%
\begin{figure*}
    \centering
    \begin{minipage}[t]{.45\textwidth}
	\includegraphics[width = .9\columnwidth]{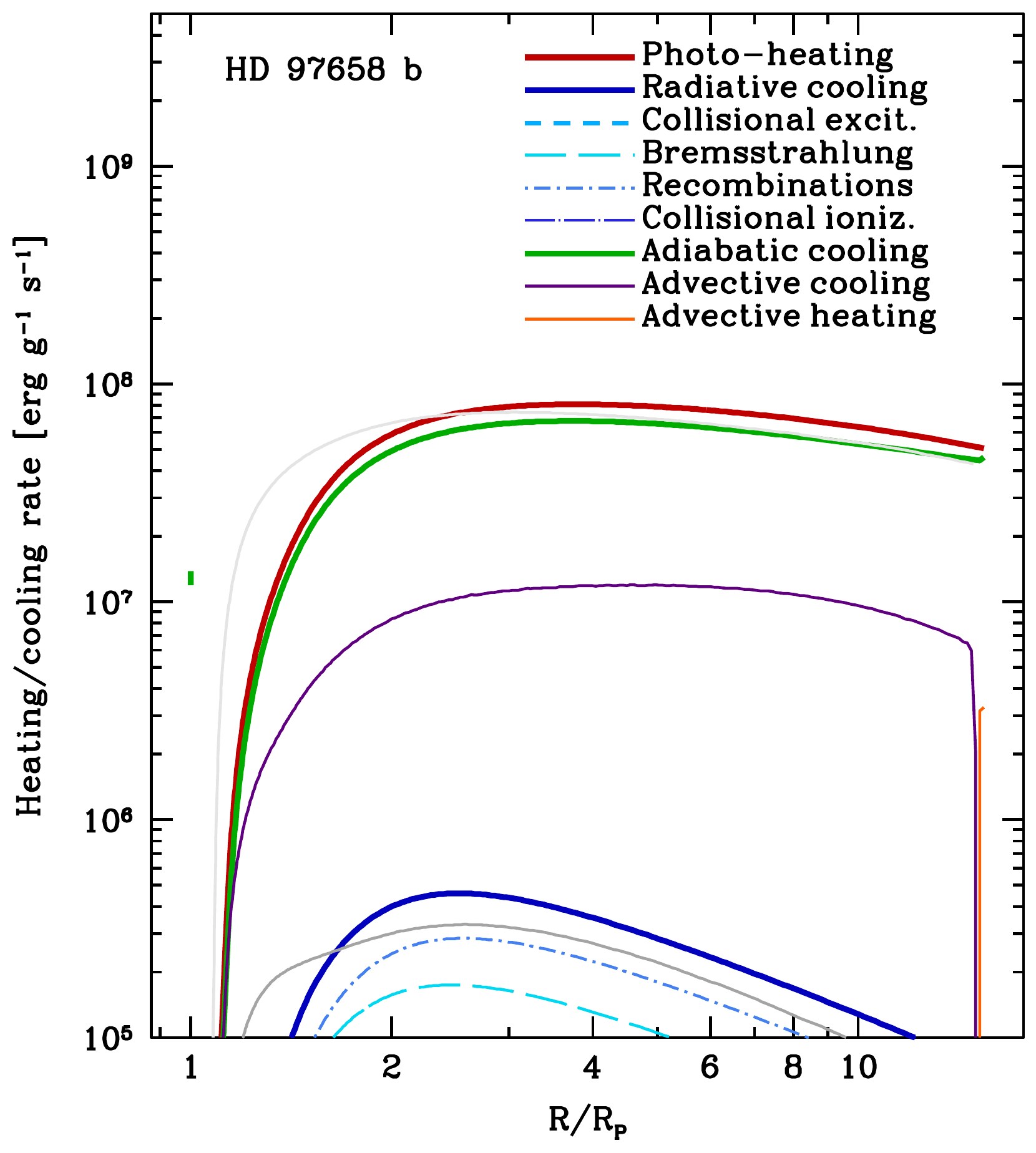}
		%\vspace{-2.5cm}
	\caption{Simulated specific heating and cooling rate profiles for the atmosphere of HD~97658~b, broken into individual, contributing mechanism (see legend for details); the dark and light grey curves, respectively, trace the photo-heating and cooling rates obtained by TPCI (S16). In the case of the low-gravity, low-irradiation planet HD~97658~b, atmospheric escape is driven by adiabatic expansion, with negligible contribution from radiative cooling. }
	\label{fig:hd-hc}
	\end{minipage}
	\hspace{.75cm}
%\end{figure}
%%%%%%%%%%%%%%%%%%%%%%%%%%%%%%%%%%%
%\begin{figure}
    \begin{minipage}[t]{.45\textwidth}
	\includegraphics[width = .9\columnwidth]{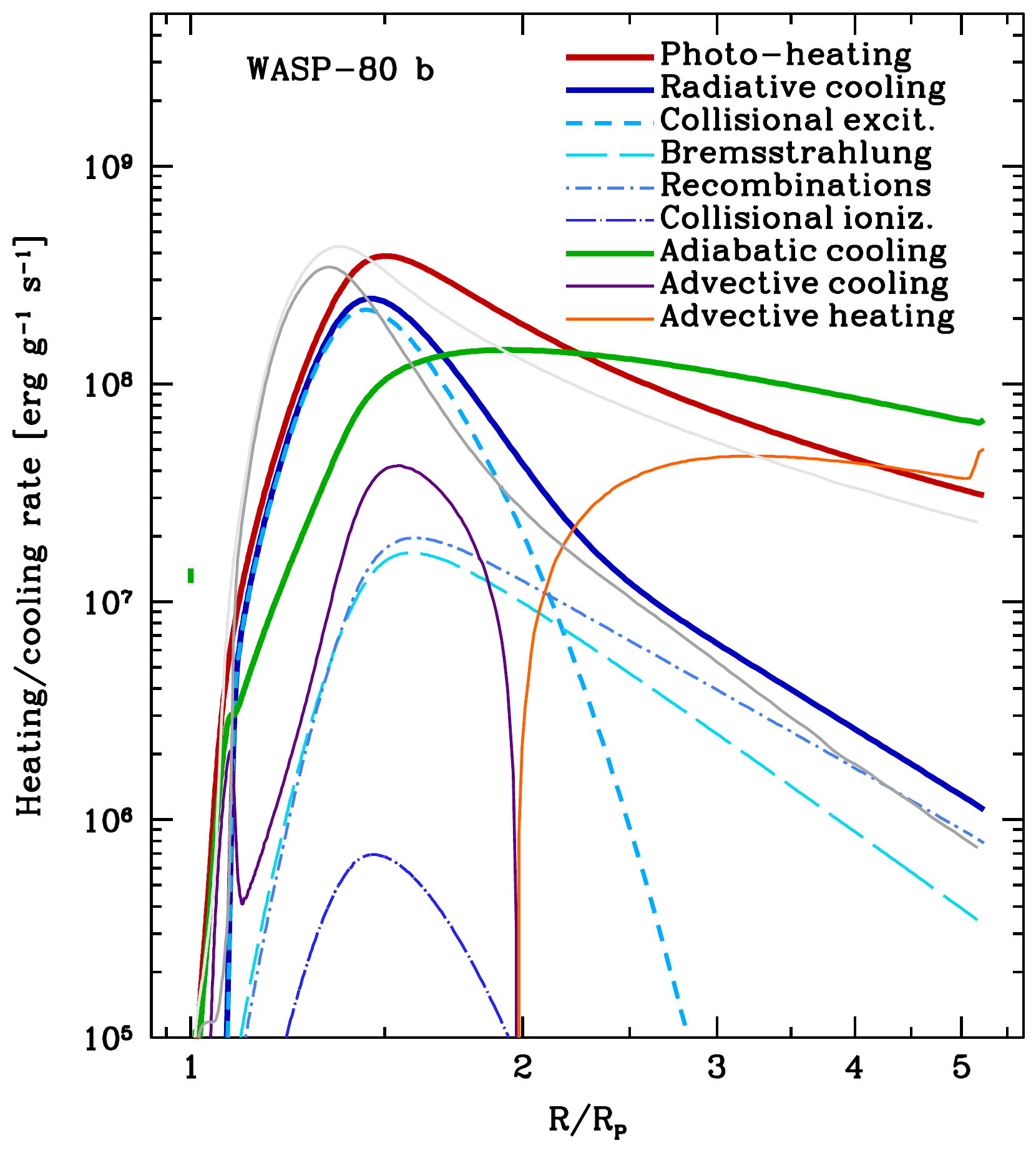}
		%\vspace{-2.5cm}
	\caption{Same as for Figure \ref{fig:hd-hc}, but for WASP-80~b. In the case of this moderate-gravity, moderate-irradiation planet, atmospheric escape is mainly driven by radiative cooling in the innermost region, albeit with a non-negligible contribution from adiabatic expansion; further out, where the ionization fraction approaches 100\%, adiabatic expansion takes over and dominates the cooling.}
	\label{fig:wasp80-hc}
	\end{minipage}
\end{figure*}
%%%%%%%%%%%%%%%%%%%%%%%%%%%%%%%%%%%
\begin{figure*}
 \centering
    \begin{minipage}[t]{.45\textwidth}
	\includegraphics[width = .9\columnwidth]{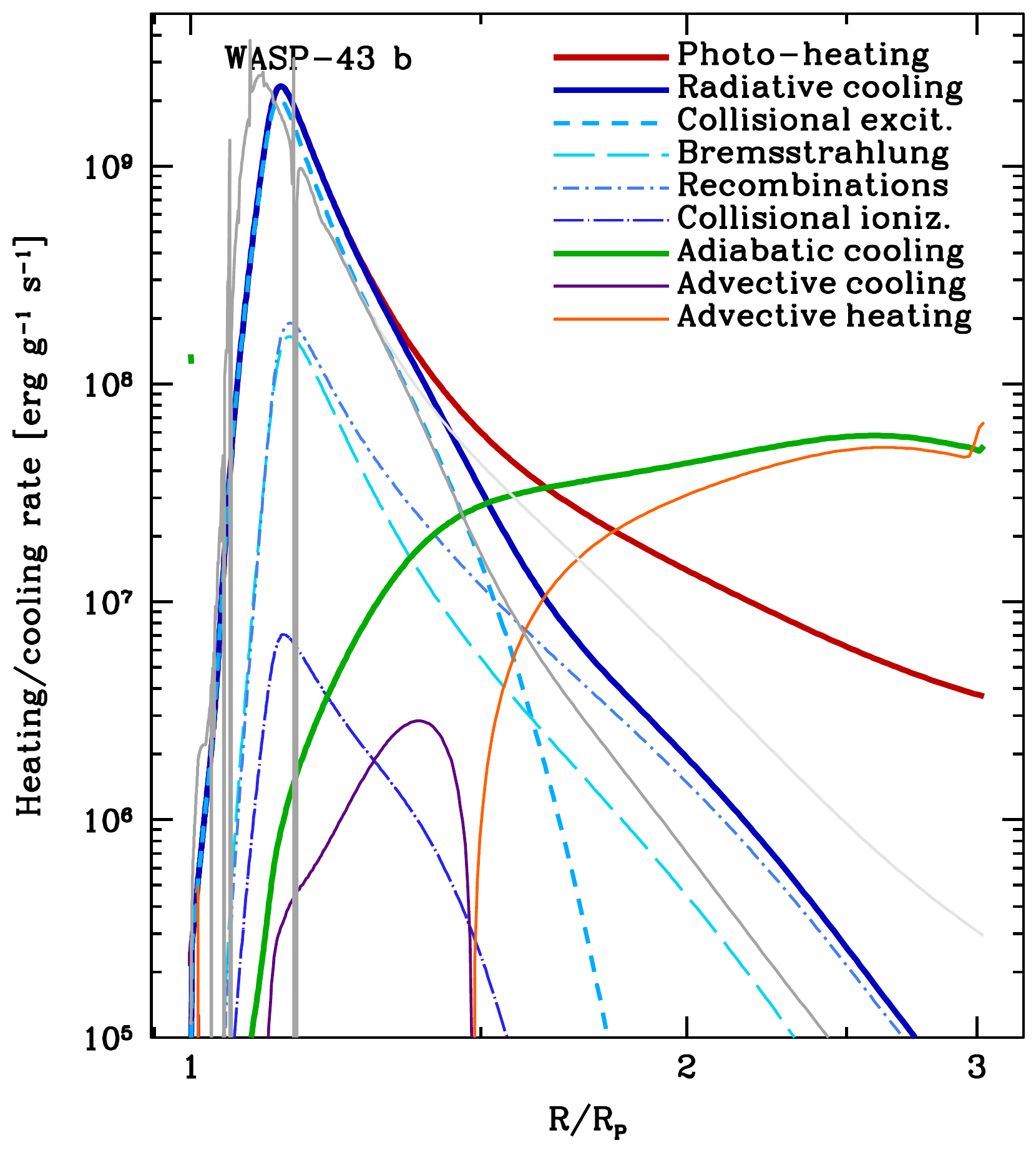}
		%\vspace{-2.5cm}
	\caption{Same as for Figure \ref{fig:hd-hc}, but for WASP-43~b. In the case of this high-gravity, high-irradiation planet, atmospheric escape is entirely driven by radiative cooling in the innermost region; only further out, where the ionization fraction approaches 100\%, adiabatic expansion takes over and dominates the cooling. }
	\label{fig:wasp43-hc}
	\end{minipage}
	\hspace{.75cm}
%\end{figure}
%
%\begin{figure}
    \begin{minipage}[t]{.45\textwidth}
	\includegraphics[width = .9\columnwidth]{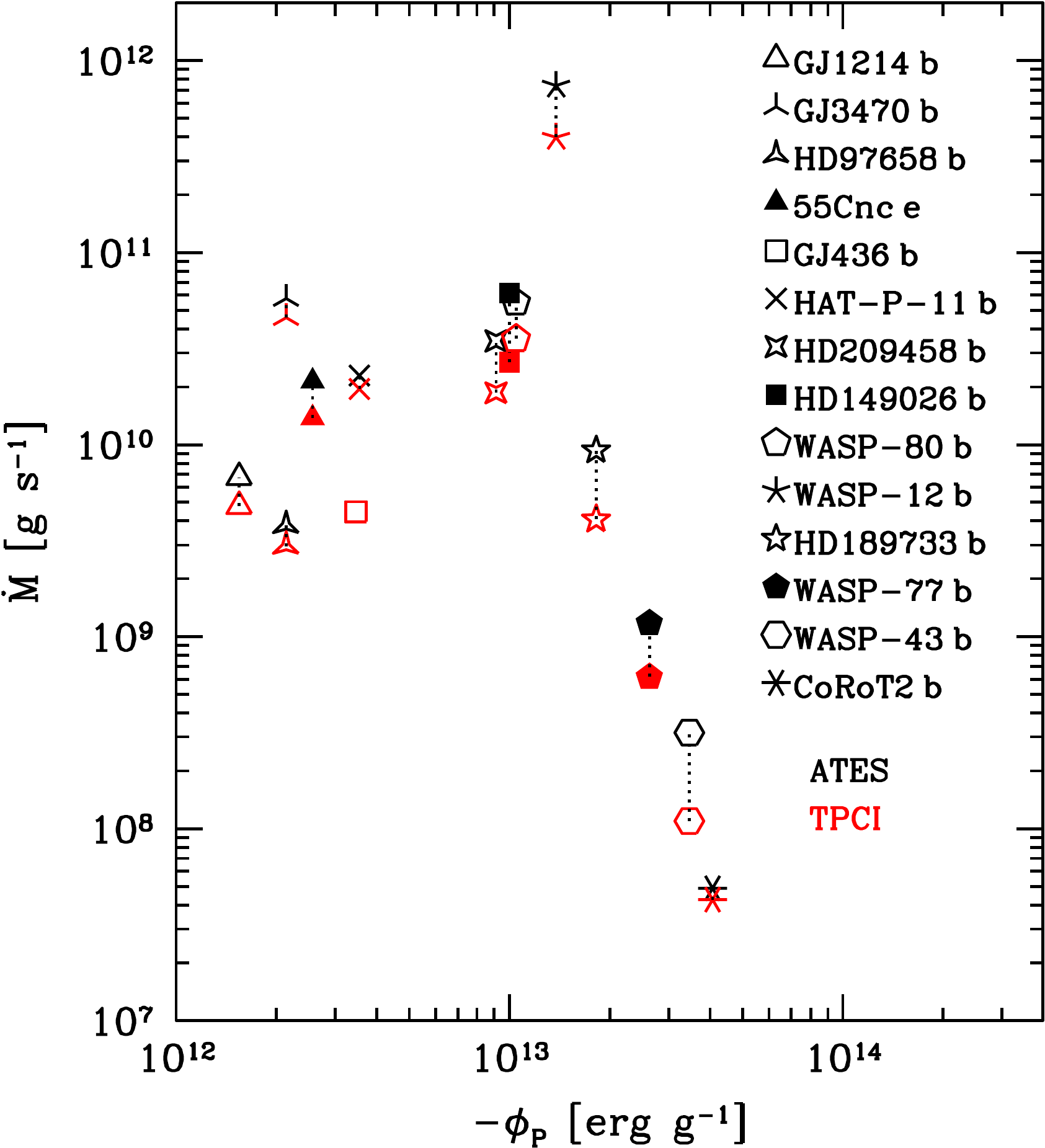}
	%\vspace{-2.5cm}
	\caption{Simulated, steady-state mass outflow rates from the 14 planetary systems considered in this work. The values obtained by ATES (in black) are compared to those by TPCI (S16, in red). Overall, the estimated mass loss rates agree to within a factor of 2 in all cases. In those cases where ATES differs from TPCI, the differences are always upward, and are likely due to inclusion of ion advection, which ATES implements in post-processing, yielding systematically higher atmospheric density profiles; this implementation choice, however, is chiefly responsible for the dramatic gain in computational time.}
	\label{fig:mdot}
	\end{minipage}
\end{figure*}
%%%%%%%%%%%%%%%%%%%%%%%%%%%%%%%%%%%%%%%

%%%%%%%%%%%%%%%%%%%%%%%%%%%
\begin{figure}
    \includegraphics[width = 1. \columnwidth]{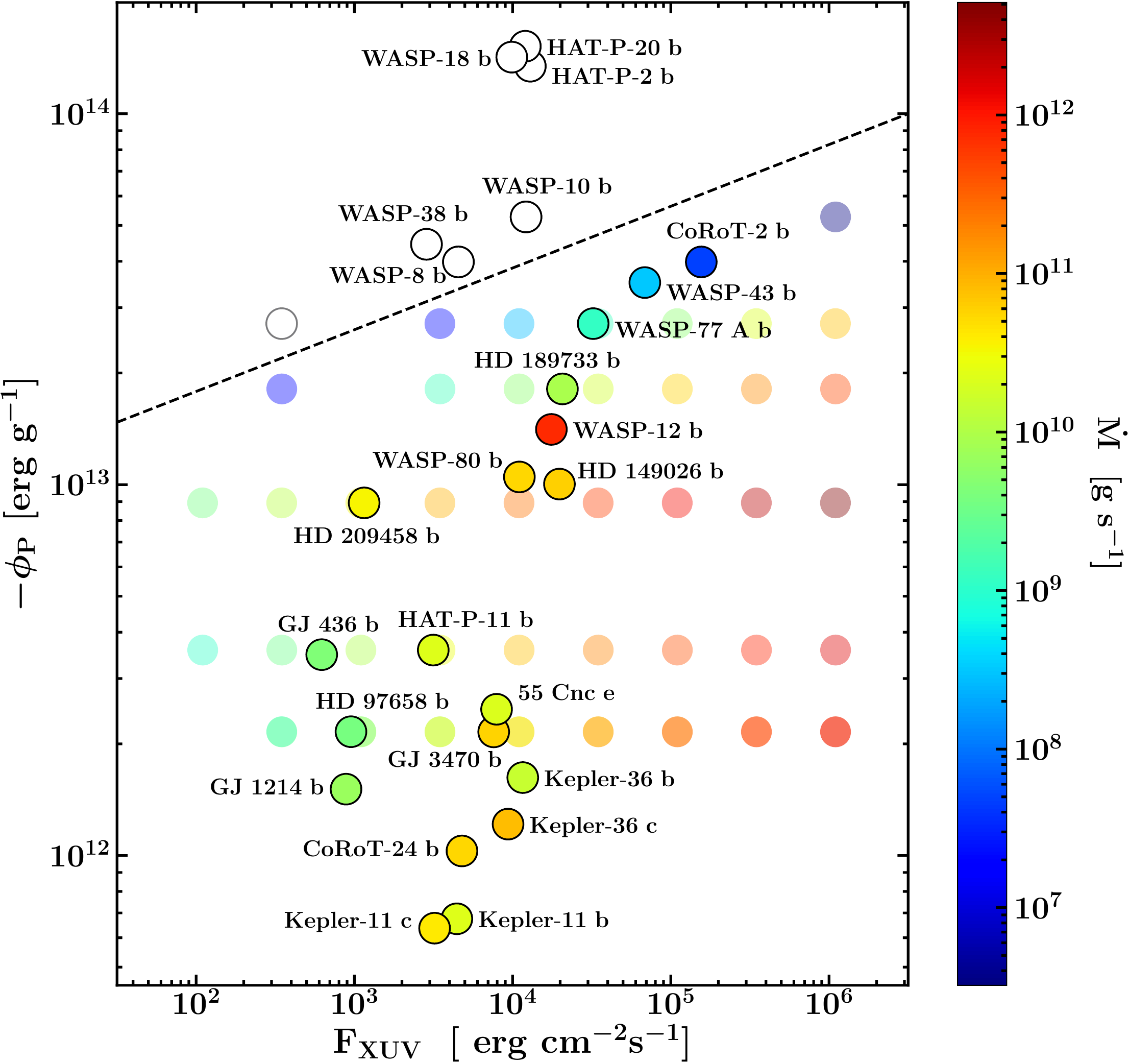}
	\caption{All the planetary systems simulated in this study are shown in the ($\phi_p$:$F_{\rm XUV}$) plane, and color-coded based on the estimated mass loss rate illustrated by the right-hand side bar. Open, color-less circles show those cases for which ATES fails to converge. In general, known planets are identified by their names and highlighted by a circle; all the others are ``synthetic''. ATES recovers stable, steady-state solutions for systems with $\log(-\phi_p) \lesssim 12.9 + 0.17\log F_{\rm XUV}$ (in cgs units); such convergence threshold is marked by the dashed black line. 	}
\label{fig:range}
\end{figure}
%%%%%%%%%%%%%%%%%%%%%%%%%%%

\section{Results and discussion}
\label{sec:Results}

A thorough discussion of the properties of the simulated atmospheres on a case-by-case basis can also be found in S16; here, we focus on presenting the results of our numerical simulations vis-à-vis those obtained by S16 with TPCI (\citealt{Salz2015}). Specifically, S16 focus on a sample of 18 nearby (within 120 pc) planets, selected a priori on the basis of the expected detectability of their out-flowing atmospheres through Ly${\alpha}$ transit spectroscopy; 14 out of those are actually found by TPCI to have non-negligible outflow rates\footnote{We caution that, in the case of WASP-12~b and HD~209458~b, the simulated (both by TPCI and ATES) mass loss rates are based on existing stellar flux {\it limits}; as such, they ought to be considered as strict upper limits.}. For comparison purposes, we run ATES on the same 14 systems, and adopting the same planetary and stellar parameters (listed in Table~\ref{tab:planets}), with the only notable difference that, whereas S16 estimate the SED of the host stars using a complex piece-wise reconstruction method (see section 2.2 of S16), we take their estimated X-ray and EUV luminosities at face value, and use them to normalize a $\propto E^{-1}$ spectrum in either bands (noting that the stellar SEDs in figure 1 of S16 are roughly consistent with flat spectra in $\nu F_{\nu}$).  

This simplification is motivated by the following reasoning. In the first approximation, the problem of atmospheric photo-heating can be thought of as an inverse Str\"omgren sphere.
As for the textbook case, where one only distinguishes between fully neutral and fully ionized regions, the location and velocity of the ionization front depend only on the rate of photons with energies above a given element ionization threshold. This is true even for the more realistic case when the photon rate is weighted over the appropriate (frequency-dependent) cross sections. Thus, as long as the underlying spectral shape and normalization preserve the total number of photons per unit time in each band, the properties of the ionization front are fairly insensitive to small features in the stellar SED. \\

Ultimately, our approach is validated by the remarkably good agreement between the outflow parameters estimated by ATES and those obtained by TPCI; the density, velocity, temperature, pressure and ionization profiles obtained from both codes are shown in Figures~\ref{fig:hd-six}, \ref{fig:wasp80-six}, and \ref{fig:wasp43-six} for three case studies: HD~97658~b (low gravity, low irradiation), WASP-80~b (moderate gravity, moderate irradiation), and WASP-43~b (high gravity, high irradiation), respectively. 

As noted by S16, the simulated atmospheres have qualitatively similar structures; the temperature profiles exhibit a very sharp rise starting from the height where the bulk of the stellar radiation is absorbed. This happens right at the planet radius for the case of WASP-80~b and WASP-43~b, whereas the temperature begins to rise sharply at $\simeq 1.1$ $R_P$ in the case of HD~97658~b, which has a higher atmospheric density.
The steep temperature gradient of the lower atmosphere is responsible for driving the atmospheric expansion against the gravitational pull of the planet. The temperature profiles all reach a maximum value further out, where the net cooling rate starts to decrease (see below). 
Mass outflows are initially very slow, with inferred velocities of a few cm s$^{-1}$ at the inner boundary, and reach supersonic velocities at the Roche lobe height, with steady-state values between 10 and 20 km s$^{-1}$. 

In general, sharp ionization fronts are typical of highly irradiated planets (such as WASP-43~b), whereas more gradual temperature profiles are typical of less irradiated planets with shallow ionization fronts.
The structure and composition of the atmosphere depend on the strength of the stellar irradiation: WASP-43~b, with an incident EUV flux $\log{F_{\text{EUV}}}=4.82$ \citep[in cgs units, i.e., close to 10,000 the solar irradiance at Earth in the same band; see, e.g.,][]{Ribas2005}, exhibits a very sharp hydrogen ionization front (Figure~\ref{fig:wasp43-six}, lower panels). A thin layer of neutral hydrogen (\HI) is confined within less than 1.3 planetary radii, above which hydrogen is fully ionized (the \HI\ fraction is $<10$\% at $1.2~R_p$). Unsurprisingly, the neutral helium profile (\HeI) tracks that of \HI, whereas the percentage of double-ionized helium (\HeIII), which is higher than 90\% beyond $2~R_p$, starts to drop below this height, and reaches $10$\%\ at $1.2~R_p$, where the \HeII\ fraction peaks (reaching $\simeq 80$\%). By comparison, HD~97658~b experiences a factor 70 lower stellar irradiation compared to WASP-43~b; its ionization front is very shallow (Figure~\ref{fig:hd-six}); the \HI\ fraction declines gently, from $100$\% at $R_p$ to $\sim 40$\% at the Roche height; the \HeI\ profile mirrors \HI's qualitatively, whereas the percentage of \HeII\ increases from close to zero at the inner boundary to about $40$\% at the Roche height, with \HeIII\ never reaching above a few per cent. The ionization profiles of WASP-80~b--whose stellar irradiance is a factor 6 lower than WASP-80~b, and a factor 11 higher than HD~97658~b--are qualitatively intermediate (see Figure~\ref{fig:wasp80-six}). 

The outflow behavior can be better understood by examining the contributions to the atmospheric heating and cooling from different mechanisms; these are shown in Figures ~\ref{fig:hd-hc}, \ref{fig:wasp80-hc}, and \ref{fig:wasp43-hc} for the same three planets discussed above. The stellar photo-heating rate is represented by the solid, thick red line; the radiative cooling term, represented by the thick, solid blue line, is given by the sum of all the possible contributions: collisional excitation (i.e., Ly${\alpha}$), bremsstrahlung, recombinations and collisional ionization (shown with different line styles and shades of blue); adiabatic cooling is represented by the solid thick green line, whereas advective cooling and heating are shown as solid purple and orange lines, respectively. For comparison, the radiative heating and cooling terms from TPCI are shown as light and dark grey lines, respectively. 
The most notable difference in the heating/cooling rates amongst the three planets has to do with the relative importance of radiative vs. adiabatic cooling. Whereas the latter completely dominates over the former across the entire domain for HD~97658~b, the situation is nearly reversed for the high-gravity/irradiation planet WASP-43~b, where radiative (and primarily Ly${\alpha}$) cooling within $\simlt 1.5~R_P$ exceeds adiabatic cooling by up to 2 orders of magnitude; further out, the high ionization fraction makes radiative cooling inefficient. 
Once again, the behavior of WASP-80~b is intermediate between HD~97658~b and WASP-43~b, in that radiative cooling here exceeds adiabatic cooling in the inner regions, albeit not as strikingly as in the case of WASP-43~b. 

Last, in Figure~\ref{fig:mdot} we compare the steady-state mass outflows estimated by ATES (in black) vs. TPCI (in red). Overall, the agreement is very good, to within a factor of 2, which is arguably lower than the ``systematic'' uncertainties associated with the modeling (see \S~3.2 in S16 for quantitative estimates). 
Whenever they differ, the ATES mass outflow rates tend to be higher than TPCI's. A close inspection of the outflow properties suggests that this difference is rooted in the density profiles, as ATES' are slightly higher than TPCI's (this is true for all the simulated systems). Although the impact on the resulting mass outflow rates is modulated by the velocity profiles, for which we find no systematic trend, higher density profiles are bound to yield higher steady-state mass outflow rates.
Considering that ATES employs the same boundary conditions as TPCI at the planet radius, the higher densities are likely to arise from the fact that ATES implements ion advection in post-processing (\S~\ref{sec:ModelDescription}) as opposed to at each time-step. In fact, ion advection has the effect to alter (ever so slightly) the atmospheric density at each time-step; in turn, this yields slightly different heating rates, and thus dynamical evolution. By accounting for the effects of advection in post-processing, ATES is likely to underestimate such time-integrated advection effects. At the same time, it is important to stress that doing so massively reduces the computational time whilst recovering realistic ionization fraction profiles (see Figure~\ref{fig:IonAdvection}). \\

As stated in \S~\ref{subsec:ModelHydro} , ATES does not include any molecular forms of hydrogen. According to \cite{odert20}, the role of H molecules is likely to be significant only for cool(er) atmospheres, and/or very close to the planet, at $\lesssim 1.1 R_p$. More specifically, the main molecular coolant in H$_2$ dominated atmospheres is IR radiation from H$_3^+$; such cooling is expected to be negligible at small orbital distances or high EUV fluxes (\citealt{Koskinen2007, Shaikhislamov2014, Chadney2015}; however, see, e.g., \citealt{Shematovich2010} for the role of H$_2$ dissociation in producing supra-thermal H atoms). Whereas we do not expect that the omission of H molecules has any significant impact on the results presented here, i.e., for moderately and highly irradiated planets ($F_{\text{XUV}}\gtrsim 10^2-10^3$ in cgs units), we plan to include molecular hydrogen in the next release of the code.\\
 
We now turn our attention to the treatment of 2D effects. As discussed in \S\ref{subsec:2Deffects}, different methods are employed in the literature to account for the fact that the photoionizing photons see different optical depths through the atmosphere, as well as averaging the photo-heating rate over the planet day-side. In order to allow for a proper comparison with TPCI, the simulations presented here were run by adopting the same prescription as S16, i.e., by diving the output mass loss rate by 4 (method (i)). In order to illustrate the effects of different choices--we re-run ATES using  (ii) the prescription by \cite{odert20}, as well as (iii) dividing the photo-heating by a factor 4 (which we suggest would be appropriate for the case of a rotating planet), and (iv) dividing both the photo-heating rate and the output mass outflow rate by 2 (appropriate for the case of a a tidally locked planet). 

For the case of GJ~3470~b, the resulting mass outflow rates differ by a factor 2 at most; specifically, we obtain $\log{\dot{M}}=10.76$ using method (i), i.e. the same method as S16 (vs. $10.66$ actually reported by S16). By comparison, method (ii) yields $\log{\dot{M}}=10.93$; method (iii) yields  $\log{\dot{M}}=10.81$, whereas method (iv) yields $\log{\dot{M}}=10.86$. We expect that the magnitude of the difference will be greater for planets where the atmospheric cooling is dominated by radiative (as opposed to adiabatic) cooling; in this respect, GJ~3470~b can be thought of as an intermediate case, where both cooling channels contribute equally. We verified that this is indeed the case by carrying out the same set of simulations for the highly irradiated gas giant WASP-77~A~b; in this case, the resulting mass outflow rates obtained by ATES are $\log{\dot{M}}=9.07, 9.52, 8.88, 8.98$ for method (i), (ii), (iii) and (iv), respectively, i.e., a factor $\sim 4$ difference at most. We caution, however, that the highest $\dot{M}$ results from 
dividing the stellar flux by a factor $(1+\alpha\tau_E)$, with $\alpha=4$ \citep{odert20}; this choice of $\alpha$ is meant to approximate the averaged 2D case in the specific case of HD~189733~b, and it is thus not obvious whether it can be extended to the case of a planet which is likely to have a more extended atmosphere.

\subsection{ATES: Applicability and numerical limitations}
ATES' range of applicability and/or validity is bounded by two classes of limitations: physical and numerical. 
The former include the 1-D approximation, the lack of H molecules, and neglecting ion-ion interactions, conductivity and the mixing of different species. The effects and implications of these approximations are discussed at various points throughout the Paper. Here, we focus on the latter; specifically, we aim to define ATES' range of applicability by performing a numerical convergence analysis on a physically relevant, bounded subspace of the $F_{\rm XUV}$:$\phi_p$ parameter space (planetary stellar flux and gravitational energy, respectively).
Having demonstrated that ATES yields good agreement with the same planets for which S16 are able to obtain steady-state mass outflows, we test the code under more extreme choices of input parameters, i.e, for extremely irradiated atmospheres and for quasi-stable atmospheres (such as can be expected for gas giants that experience moderate to low irradiation).\\
To this end, we start with simulating a handful of additional, well-known systems: five low-mass planets whose atmospheres are expected to be undergoing ``boil-off'' \citep{owenwu16}, i.e.: CoRoT-24~b, Kepler-36~b,c, and Kepler-11~b,c (see, e.g., \citealt{Lammer2016,Owen2016,Cubillos2017})\footnote{The planetary parameters for these systems were taken from \url{exoplanet.eu}, whereas the stellar luminosities were estimated following \citet{Lammer2016}}
To those we add new simulations for a handful of moderately-irradiated gas giants whose atmospheres are likely not undergoing hydro-dynamical escape, namely: WASP-38~b, WASP-8~b, WASP-10~b, WASP-18~b, HAT-P-2~b, and HAT-P-20~b; both stellar and planetary parameters for these systems were taken from S16 and references therein. 
The results of this investigation are summarized in Figure~\ref{fig:range}. Those systems for which ATES reaches/fails to reach convergence are represented as filled and open circles, respectively. For the former group, decreasing values of mass outflow rates are characterized by progressively cooler colors. In general, we observe a trend whereby the code fails to reach convergence for the heaviest gas giants. The lack of known gas giants experiencing low irradiation (namely, with $\log(-\phi_p)\simgt 13$ and $\log F_{\rm XUV}\simlt 3$, in cgs units) prevents us from characterizing any possible dependence of the convergence on irradiation, although the very fact that solution convergence is reached for CoRoT-2~b and not for WASP-8~b (which have comparable $\phi_p$) suggests that irradiation does indeed play a role. 
To fully parse ATES' range of applicability, we further simulate a large set of mock planets, sampling the gravity-stellar flux plane at higher resolution. This enables us to define a quantitative criterion for convergence, which is illustrated by the dashed line in Figure~\ref{fig:range}: $\log(-\phi_p) \lesssim 12.9 + 0.17\log F_{\rm XUV}$ (cgs units). Below this approximate threshold, ATES can be reliably expected to reach convergence and yield a steady-state atmospheric mass outflow rate.
Conversely, systems above this threshold were all identified by S16 as likely undergoing Jeans escape.  

\subsection{Comparison to other existing codes}
Beside TPCI \citep{Salz2015}, which we used as a benchmark for our simulations, several other codes exist in the literature which are used to model atmospheric escape in exoplanets. 
Generally speaking, they can be divided in two categories: dedicated, proprietary hydrodynamic/radiative codes (mostly 1-D), and public, multi-purpose 3-D codes, or adaptations thereof \citep[such as TPCI,][]{Salz2015}. We note that we are intentionally limiting the discussion below to numerical works that were developed over the last 5 years, thus omitting early, seminal works. 
Notable amongst the former is the 1-D code developed by \citet{Erkaev2016}; similar to ATES in its hydrodynamics/energy and ionization balance treatment (unlike ATES, it includes thermal conduction, but neglects \He), this code has been extensively used (by the many contributing authors) to estimate the atmospheric profiles and mass outflow rates for several, well-known, highly irradiated exoplanets \citep[e.g.,][to name a few]{Erkaev2017,Fossati2017,kuby18a,odert20}. 
More recently developed (proprietary) 1-D hydrodynamics codes include \cite{Vidotto2020} (following \citealt{AllanVidotto2019}), which computes the effects of the stellar wind ram pressure on the atmospheric outflow, and \citet{Bisikalo2018}, which adapts the code by \citep{Ionov2017} to account for the role of supra-thermal photo-electrons during stellar flares.  
Somewhat separately, \cite{chenrogers16} developed a prescription to adapt the capabilities of the Modules for Experiments in Stellar Astrophysics (MESA) to model sub-Neptune-sized planets with H/He envelopes. This planetary evolution module, along with subsequent variations and/or improvements, has been widely used to ascertain the role of thermal evolution vs. stellar irradiation in shaping the observed distribution of exoplanet radii \citep{fulton17}. As an example, \cite{kuby20} and \cite{kuby21} study the evolution of planetary atmospheres under the combined effects of atmospheric mass loss (modelled using the proprietary code by \citealt{kuby18a}) and planetary thermal evolution---modelled using MESA, after \citealt{paxton19}). \\
A separate mention goes to \citet{Koskinen2014} and \citet{Khodachenko2019}, and references therein, who perform detailed 3-D hydrodynamic simulations of a handful of giant planets with proprietary codes that feature extensive chemical networks.\\
Turning to publicly available, multi-purpose codes, \citet{Debrecht2019} carry out 3-D simulations of atmospheric outflows from synthetic planets with ASTROBEAR \footnote{\url{https://www.pas.rochester.edu/astrobear}}, a parallelized, magneto-hydrodymamics (MHD) code designed for 2D and 3D adaptive mesh refinement simulations.  The same code is used by \citet{Debrecht2020} to investigate the role of the stellar Ly$\alpha$ radiation pressure on the outflow dynamics. In these studies ASTROBEAR is set up to simulate a single frequency, planar radiation field impacting on a primordial, atomic H atmosphere.
Along the same lines, \citet{McCann2019} adopt the 3-D, MHD Eulerian code ATHENA \citep{Stone2008}, complemented by dedicated ionization and radiative transfer modules, to simulate planetary atmospheres, also including the stellar wind and Coriolis force effects. A similar approach is that by \citet{Esquivel2019} who adapt the publicly available 3-D MHD code GUACHO \citep{guacho} to investigate the interaction between the stellar and planetary winds, also accounting for radiation pressure and charge exchange.\\
The latter type of studies are typically focused on a specific planetary system with a wealth of available data, where employing a sophisticated 3-D code allows for a detailed (if time-consuming) morphological characterization of the outflow, including, e.g., the possible development of cometary tails. In contrast, 1-D codes that allow for a swift estimate of the outflow properties, including the instantaneous mass-outflow rate, can be thought of as an efficient tool for assessing the most promising targets for intensive spectroscopic follow-ups out of a large pool of systems. Amongst these, ATES has the added benefit of being fast--a typical run takes minutes-to-hours on a standard laptop to achieve convergence--and publicly available.

\section{Conclusions}
In summary, we have developed a new and efficient photoionization hydrodynamics code that can be easily employed to readily estimate the instantaneous atmospheric mass loss rates from highly irradiated planets. The code, which is publicly available, can be run through an intuitive graphic interface where the user can specify the grid and reconstruction method of choice. 
For a given choice of planetary and stellar parameters, the code calculates the corresponding atmospheric temperature, density, velocity and ionization fraction profiles, for a primordial composition of atomic \Hy and \He, where the (user-specified) \Hy-to-\He ratio is kept fixed throughout the simulation. 

The ATES results are in very good agreement with those obtained by TPCI (The Pluto Cloudy Interface; S16) for 14 moderately-to-highly irradiated systems (see Table~\ref{tab:rates} and Figure~\ref{fig:mdot}); minor differences are likely due to the implementation of ion advection, which ATES carries out in post-processing (\S~\ref{sec:adv}). This scheme, however, results in a major speed-up in calculation, with the considered 14 systems taking between 3 minutes to up to 2 hours (depending on the actual case) on a standard 2.7 GHz quad core CPU.

A comprehensive description of the code installation and usage is also available as part of the online repository. Future developments will involve the inclusion of molecular hydrogen as well as metal cooling, the possibility to model time-variable stellar irradiation, the implementation of two-dimensional effects, and the creation of a separate module to calculate the expected Ly$\alpha$ line profiles, along with other atomic transitions that could be targeted from the ground with upcoming 30-m class facilities.

%%%%%%%%%%%%%%%%%%% ACKNOWLEDGMENTS %%%%%%%%%%%%%%%%%%

\begin{acknowledgements}
We are grateful to the anonymous reviewer for insightful and constructive comments which greatly benefited the paper.
\end{acknowledgements}

%%%%%%%%%%%%%%%%%%%% REFERENCES %%%%%%%%%%%%%%%%%%
\clearpage
\bibliographystyle{aa} % style aa.bst
\bibliography{bibliography.bib} % your references Yourfile.bib

%%%%%%%%%%%%%%%%%% APPENDICES %%%%%%%%%%%%%%%%%%%%%

\begin{appendix} %First appendix
\section{Rates}
\label{app:rates}

In Table~\ref{tab:cool} we report the cooling rates used in our code, while Table~\ref{tab:rates} shows recombination and collisional ionization coefficients adopted in the ionization equilibrium calculation.

%%%%%%%%%%%%%%%%%%%%%%%%%%%%%%%%%%%%%%%%%%%%%%%%%%%%
\begin{table*}
\renewcommand{\arraystretch}{1.3}
\centering
\caption{Cooling rates per free electron $\Lambda$. $T$ is temperature in [K], while $Z_i$ the atomic number of the $i-$th ion.}
\label{tab:cool}
\makebox[\textwidth][c]{
\begin{tabular}{clll}
\hline
\hline 
\textbf{Process} & \textbf{Species} & \textbf{$\Lambda$ [erg/s]} & \textbf{Ref.}\\
\hline
\hline
\noalign{\vskip 3pt} 

Bremsstrahlung	&		&	$1.426\cdot 10^{-27}Z_i^2\sqrt{T}~G_F(Z_i,T)n_i$   $[T<3.2\cdot 10^5Z_i^2]$  & (2) \Tstrut\\
				&		&	$G_F(Z_i,T)=0.79464+0.1243\log_{10}(T/Z_i^2)$  \Bstrut\\
%-------------------------------------------------------------------------
\noalign{\vskip 3pt} 
\hline
\noalign{\vskip 3pt} 
					   & \HI   & $7.5\cdot 10^{-19}\left(1+\sqrt{\dfrac{T}{10^5}}\right)^{-1}\exp\left(-\dfrac{118348}{T}\right)~n_\HI$ & (2) \Tstrut\\
Collisional excitation & \HeI~($1^1S$)  & $1.1\cdot 10^{-19}T^{0.082}\exp\left(-\dfrac{230000}{T}\right)~n_\HeI$ & (2)\\
					   & \HeII & $5.54\cdot 10^{-17}T^{-0.397}\left(1+\sqrt{\dfrac{T}{10^5}}\right)^{-1}\exp\left(-\dfrac{473638}{T}\right)~n_\HeII$ & (2) \Bstrut\\
\noalign{\vskip 3pt} 
%-------------------------------------------------------------
\hline
\noalign{\vskip 3pt} 
				&	\HII    &	 $3.435\cdot 10^{-30}T\left(\dfrac{315614}{T}\right)^{1.970}\left[1+\left(\dfrac{140273}{T}\right)^{0.376}\right]^{-3.720}~n_\HII$ & (1)  \Tstrut \\
Recombination	&	\HeII   &	 $1.38\cdot 10^{-16}T\alpha^\text{rec}_{\HeII} ~n_\HeII$ & (1)\\
				&	\HeIII  &	 $2.748\cdot 10^{-29}T\left(\dfrac{1263030}{T}\right)^{1.970}\left[1+\left(\dfrac{561347}{T}\right)^{0.376}\right]^{-3.720}~n_\HeIII$  & (1) \Bstrut \\

%-----------------------------------------------------------------------------------------------------------
\noalign{\vskip 3pt} 
\hline
\noalign{\vskip 3pt} 
                        	& \HI   &	$2.179\cdot 10^{-11}\alpha^\text{ion}_{\HI}~n_\HI$     & (2) \Tstrut \\
Collisional ionization		& \HeI  &	$3.940\cdot 10^{-11}\alpha^\text{ion}_{\HeI}~n_\HeI$   & (2)         \\
						    & \HeII &   $8.715\cdot 10^{-11}\alpha^\text{ion}_{\HeII}~n_\HeII$ & (1) \Bstrut \\
\noalign{\vskip 3pt} 
\noalign{\vskip 3pt}
\hline
\end{tabular}}
\tablebib{(1)~\citet{Hui1997}; (2) \citet{Glover2007}.}
\end{table*}
%%%%%%%%%%%%%%%%%%%%%%%%%%%%%%%%%%%%%%%%%%%%%%%%%%%%%%%%%%%%%%%

%%%%%%%%%%%%%%%%%%%%%%%%%%%%%%%%%%%%%%%%%%%%%%%%%%%%%%%%%%%%%%%%
\begin{table*}
\renewcommand{\arraystretch}{1.3}
\centering
\caption{Recombination and collisional rate coefficients adopted. In the rates, $T$ is the temperature in [K], while $\theta_e\equiv \ln{(kT/[\rm{eV}])}$.}
\label{tab:rates}
\makebox[\textwidth][c]{
\begin{tabular}{clll}

\hline 
\hline
\noalign{\vskip 3pt} 
\textbf{Reaction} & \textbf{Symbol} & \textbf{Rate coefficient [cm$\mathbf{^3}$~s$\mathbf{^{-1}}$]} & \textbf{Ref.}\Tstrut\Bstrut\\
\noalign{\vskip 3pt} 
\hline 
\hline
\noalign{\vskip 3pt}
$\HII +e^-\to \HI +\gamma$ & $\alpha^\text{rec}_{\HII}$ & $2.753\cdot 10^{-14}\left(\dfrac{315614}{T}\right)^{1.5}\left[1+\left(\dfrac{115188}{T}\right)^{0.407}\right]^{-2.242}$ & (1) \Tstrut\\
%-------------------------
$\HeII +e^-\to \HeI+\gamma$ & $\alpha^\text{rec}_{\HeII}$ & $1.26\cdot 10^{-14}\left(\dfrac{570670}{T}\right)^{0.750}$ & (1) \\
$\HeIII +e^-\to \HeII +\gamma$ & $\alpha^\text{rec}_{\HeIII }$ & $5.506\cdot 10^{-14}\left(\dfrac{1263030}{T}\right)^{1.5}\left[1+\left(\dfrac{460960}{T}\right)^{0.407}\right]^{-2.242}$ & (1)\Tstrut\\
%-------------------------
\noalign{\vskip 3pt}
\hline
\noalign{\vskip 3pt}
$\HI +e^-\to \HII +e^-+e^-$ & $\alpha^\text{ion}_{\HI}$ & $\exp (-3.271396786\cdot 10^1+1.35365560\cdot 10^1\theta_e$ & (2) \Tstrut\\
						&					& $\quad~~ -5.73932875\cdot 10^0\theta_e^2+1.56315498 \cdot 10^0\theta_e^3$\\
						&					& $\quad~~ -2.87705600\cdot 10^{-1}\theta_e^4+3.48255977\cdot 10^{-2}\theta_e^5$\\
						&					& $\quad~~ -2.63197617\cdot 10^{-3}\theta_e^6+1.11954395\cdot 10^{-4}\theta_e^7$\\
						&					& $\quad~~ -2.03914985\cdot 10^{-6}\theta_e^8)$\\
%------------------------						
$\HeI +e^-\to \HeII +e^-+e^-$ & $\alpha^\text{ion}_{\HeI }$ & $\exp(-4.409864886\cdot 10^{1} + 2.391596563\cdot 10^{1}\theta_e$ & (2) \\
						  &					   & $\quad~~-1.07532302\cdot 10^{1}\theta_e^2 + 3.05803875\cdot 10^0\theta_e^3$\\
						  & 				   & $\quad~~-5.6851189\cdot 10^{-1}\theta_e^4 + 6.79539123\cdot 10^{-2}\theta_e^5$\\
						  & 				   & $\quad~~-5.0090561\cdot 10^{-3}\theta_e^6 + 2.06723616\cdot 10^{-4}\theta_e^7$\\
						  & 				   & $\quad~~-3.64916141\cdot 10^{-6}\theta_e^8)$\\
$\HeII +e^-\to \HeIII +e^-+e^-$ & $\alpha^\text{ion}_{\HeII}$ & $19.95T^{3/2}\exp\left(\dfrac{631515}{T}\right)\left(\dfrac{1263030}{T}\right)^{-1.089}\left[1+\left(\dfrac{2283960}{T}\right)^{0.735}\right]^{-1.275}$ & (1) \Bstrut\\
\noalign{\vskip 3pt}
\hline
\end{tabular}}
\tablebib{(1)~\citet{Hui1997}; (2) \citet{Glover2007}.}
\end{table*}
%%%%%%%%%%%%%%%%%%%%%%%%%%%%%%%%%%%%%%%%%%%%%%%%%%%%%%

\end{appendix}

\end{document}